
\input phyzzx.tex
\pubnum{ROM2F-94-17}
\def\a{\alpha}
\def\b{\beta}

\def\d{\delta}

\def\l{\lambda}
\def\m{\mu}
\def\n{\nu}
\def\r{\rho}
\def\s{\sigma}
\def\t{\tau}
\def\z{\zeta}
\def\mnr{\mu\nu\rho}
\def\mn{\mu\nu}

\def\wh{\widehat}
\def\wt{\widetilde}
\def\de{\partial}

\def\ddpsi{{\partial \over \partial\psi}}
\def\ddtheta{{\partial \over \partial\theta}}
\def\ddphi{{\partial \over \partial\varphi}}
\def\adot{\dot\alpha}
\def\bdot{\dot\beta}

\def\aadot{\alpha\dot\alpha}

\def\adota{\dot\alpha\alpha}

\def\adotb{\dot\alpha\beta}

\def\eps{\varepsilon^{\alpha\beta}}
\def\epsdot{\varepsilon^{\dot\alpha\dot\beta}}
\def\epsinv{\varepsilon_{\alpha\beta}}
\def\epsdotinv{\varepsilon_{\dot\alpha\dot\beta}}
\def\sigaaadot{\sigma^a{}_{\alpha\dot\alpha}}

\def\sigabbdot{\sigma^a{}_{\beta\dot\beta}}

\def\sigbabdot{\sigma^b{}_{\alpha\dot\beta}}

\def\sigaadota{\bar\sigma_a{}^{\dot\alpha\alpha}}
\def\sigabdotb{\bar\sigma_a{}^{\dot\beta\beta}}
\def\sigaadotb{\bar\sigma_a{}^{\dot\alpha\beta}}

\def\sigbadotb{\bar\sigma_b{}^{\dot\alpha\beta}}

\def\du{\scriptscriptstyle D}
\def\mezzo{{1 \over 2}}
\def\unoar{1-({a \over r})^4}
\def\ar{({a \over r})^4}
\def\ap{\alpha^\prime}
\def\unita{{1 \kern-.30em 1}}
\def\fey{{\big / \kern-.80em D}}
\def\complex{{\kern .1em {\raise .47ex \hbox {$\scriptscriptstyle |$}}
\kern -.4em {\rm C}}}
\def\zet{{Z \kern-.45em Z}}
\def\real{{\vrule height 1.6ex width 0.05em depth 0ex
\kern -0.06em {\rm R}}}
\def\rational{{\kern .1em {\raise .47ex \hbox{$\scripscriptstyle |$}}
\kern -.35em {\rm Q}}}
\titlepage
\title{ALE Instantons in String Effective Theory}
\author{M.~Bianchi, F.~Fucito, G.C.~Rossi}
\address{Dipartimento di Fisica, Universit\`a di Roma II ``Tor
Vergata''
I.N.F.N. Sezione di Roma II ``Tor Vergata'',00133 Roma, Italy}
\andauthor{ M.~Martellini}
\address{Dipartimento di Fisica, Universit\`a di Milano, 20133
Milano, Italy
Sezione I.N.F.N. dell'Universit\`a di Pavia, 27100 Pavia, Italy}

\abstract
{We show that the classical equations of motion of the low-energy effective
field theory describing the massless modes of the heterotic (or type I)
string admit two classes of supersymmetric self--dual backgrounds.
The first class, which was already considered in the literature,
consists of solutions with a (conformally) flat metric coupled to
axionic instantons.
The second includes Asymptotically Locally Euclidean (ALE) gravitational
instantonic backgrounds coupled to gauge instantons
through the so--called  ``standard embedding''. We show that some elements of
these two classes
of solutions are dual to each other in the sense of Buscher's duality.
We give a world--sheet interpretation of the heterotic ALE istanton
solutions in terms of superconformal $N=(4,4)$ $\s$--models and argue for
their validity to all orders in $\alpha^\prime$.
Specializing the gravitational background to the  Eguchi--Hanson instanton,
we compute the indices of the fermionic operators and give the explicit
form of all the relevant fermionic and bosonic zero--modes.}
\endpage
\pagenumber=1

\chapter{Introduction}
\REF\wittdue{M.B.~Green, J.H.~Schwarz and E.~Witten,
{\it Superstring Theory}, Cambridge University Press, 1987.}
\REF\cfmp{C.G.~Callan, D.~Friedan, E.J.~Martinec and M.J.~Perry,
Nucl.Phys. {\bf B262} (1985) 593.}
\REF\caltho{For a review see C.G.~Callan and L.~Thorlacius, {\it Particles,
Strings and Supernovae}, Vol.2 TASI 88, eds. A.~Jevicki and C.-I.~Tan,
World Scientific (1989).}
Several non trivial solutions of general relativity (black holes, plane
gravitational waves, cosmological metrics, \dots) have been shown to be also
classical solutions of some string theory.
This is usually proven by recognizing that the one loop
$\b$--functions of the associated non-linear $\s$--models, describing the
propagation of the string in the corresponding backgrounds, are zero.
In turn these conditions can be looked at as
the Euler-Lagrange equations
of an effective action describing the low energy interactions
of the massless modes of the string [\wittdue, \cfmp, \caltho].

\REF\khuri{R.R.~Khuri, Nucl.Phys. {\bf B387} (1992)
315.}
\REF\dghr{A.~Dabholkar, G.~Gibbons, J.~Harvey and
F.R.~Ruiz, Nucl.Phys {\bf B340} (1990) 33.}
\REF\calluno{C.~Callan, J.~Harvey and
A.~Strominger, Nucl.Phys {\bf B359}(1991) 611; ibid. {\bf B367} (1991) 60.}
\REF\penta{M.J.~Duff and J.X.~Lu, Nucl.Phys. {\bf 354}  (1991) 141;
\nextline A.~Strominger,
Nucl.Phys. {\bf B343} (1990) 167.}
As is well known,
in order to find classical solutions of a supersymmetric theory,
it is enough to set to zero the fermion fields together
with their supersymmetric variations.
This is the way in which several monopole [\khuri], soliton [\dghr]
and instanton [\calluno] configurations
have been proven to be classical solutions (known as ``pentabranes" [\penta])
of the effective supergravity theory derived from the heterotic string.
In these solutions  the metric is conformally flat.

\REF\sigmaquattro{L.~Alvarez-Gaum\'e and D.Z.~Freedman, Phys.Rev.
{\bf D15} (1980) 846; Comm. Math. Phys. {\bf 80} (1981) 443;\nextline
L.~Alvarez-Gaum\'e, Nucl.Phys. {\bf 184} (1981) 180;\nextline
L.~Alvarez-Gaum\'e and P.~Ginsparg, Comm. Math. Phys. {\bf 102}
(1985) 311;\nextline
C.~Hull, Nucl.Phys. {\bf 260} (1985) 182.}
The $\s$--models describing  the propagation of the heterotic string in
these backgrounds have  $N=(4,0)$ supersymmetry on the world--sheet.
If the gauge connection
is identified with the generalized spin connection
with torsion (``standard embedding''), the resulting $\s$--model
turns out to be left-right symmetric and the
$N=(4,0)$ supersymmetry is enhanced to $N=(4,4)$. As a consequence,
all $\alpha^\prime$ corrections to the $\b$--functions are expected to be zero
[\sigmaquattro].

In particular the instantonic solution of [\calluno] (in the limit in
which the dilaton field is taken to vanish at infinity) was proven to be an
exact solution of the heterotic string equations of motion by showing
that the corresponding $\s$--model coincides with an exactly conformal
WZW model
on the group manifold $SU(2)_k\otimes U(1)_Q$, where the level $k$ and the
background charge $Q$ are related by $Q=1/\sqrt{k+2}$. With this choice
the coefficient of the conformal anomaly turns out to be $c=6$,
as in the case of four flat supercoordinates.

\REF\bdr{E.A.~Bergshoeff and M.~de~Roo,
Nucl.Phys. {\bf B328} (1989) 439.}
\REF\candelas{P.~Candelas, G.~Horowitz, A.~Strominger and
E.~Witten, Nucl. Phis. {\bf B258} (1985) 46}
Generally speaking, all higher order $\alpha^\prime$ corrections to the
heterotic string classical solutions appear
to be proportional to $trR\wedge R-trF\wedge
F$ (or contractions thereof) [\bdr], which indeed
vanish thanks to the identification of $R$ with $F$, following
from the ``standard embedding''. This observation has been widely used
to derive consistent compactifications of the heterotic string on compact
Calabi-Yau manifolds [\candelas].

\REF\char{J.M.~Charap and M.J.~Duff, Phys.Lett.
{\bf 69B} (1977) 445.}
It is curious to note in this context that, to the extent of our
knowledge, the first example of use of the
``standard embedding'' was to generate Yang-Mills instantons
out of gravitational ones [\char] in a Yang-Mills field theory coupled
to gravity.

In this paper we will consider the possibility of promoting the solutions of
the Euclidean Einstein equations, known as gravitational instantons,
to fullfledged
solutions of the heterotic string classical equations of motion.
We will mostly concentrate our attention on
gravitational instantons which correspond to non-compact, four--dimensional
Asymptotically Locally Euclidean (ALE) manifolds
admitting Ricci-flat hyperk\"ahler metrics with self--dual curvature.

\REF\rey{S.-J.~Rey, Phys.Rev. {\bf D43} (1991) 526.}
To construct a solution of the whole set of classical equations
we notice that ALE instantons satisfy
the empty space Euclidean Einstein equations. Thus in order to couple them to
non--trivial backgrounds of the other massless bosonic fields, one must
consider field configurations with zero Euclidean stress energy tensor.
Gauge instantons and axionic instantons [\rey] are solutions of
the equations of motion in the absence of
fermionic sources, which precisely meet this requirement.

For the consistency of the solution the validity of the (modified) Bianchi
identities
has to be imposed. To fulfill this requirement one is naturally led, as a
first step, to identify the gauge and the (generalized) spin connection.
In this way the Bianchi
identities are simply reduced to the requirement that the dilaton
field must be a constant or the metric conformally flat.

\REF\buscher{T.~Buscher, Phys. Lett. {\bf 159B} (1985); {\bf 194B} (1987);
{\bf 201B} (1988).}
It is remarkable that, as we will show in sect.~2.2, the solution
in which the gravitational background is a self--dual multi--center metric
and the dilaton field is a
constant, turns out to be dual, in the sense of Buscher's duality
[\buscher], to the solution found in [\khuri],
in which, viceversa, the  gravitational background is conformally
flat and the whole non triviality lies in the form of the dilaton.
\REF\rocek{M.~Ro\v cek, C.~Ahn, K.~Schoutens and A.~Sevrin, {\it Superspace
WZW Models and Black Holes}, IASSNS-HEP-91-69, Contribution to
Workshop on Superstrings and Related Topics, Trieste, Italy, Aug. 8-9,
1991, hep-th/9110035.}
As for the solution discussed in [\calluno], its dual is the stringy black
hole solution described in [\rocek].

\REF\wittop{E.~Witten, Comm. Math. Phys. {\bf117}
(1988) 353; {\it Supersymmetric Yang--Mills Theory on a Four--Manifolds},
IASSNS-HEP-94/5.}
\REF\hp{S.W.~Hawking and
C.N.~Pope, Nucl.Phys. {\bf B146} (1978) 381.}
Many topological properties of ALE manifolds have been known for a while.
The possibility of finding new topological invariants,
through the study of
topological quantum field theories [\wittop] on such manifolds, may not forego
the precise determination of the indices of the fermionic operators
and of all the relevant bosonic and fermionic zero--modes.
The computations can be drastically
simplified by exploiting the existence of a surviving global $N=1$
right--handed supersymmetry, related to the presence of two independent
covariantly constant spinors  [\hp]. This supersymmetry allows to
relate zero--modes
of the bosonic fields to those of their fermionic superpartners. In this way
one is reduced to compute fermionic zero--modes only.

\REF\eh{T.~Eguchi and A.J.~Hanson, Ann.Phys.{\bf 120} (1979) 82.}
\REF\egh{For a review on gravitational instantons see: T.~Eguchi,
P.B.~Gilkey and A.J.~Hanson, Phys.Rep. {\bf 66} (1980) 213.}
In order to be explicit we shall take the simplest among the ALE manifolds:
the Eguchi--Hanson instanton [\eh] and we complete the solution of the
classical
equations of motion by coupling it to a gauge instanton constructed by
identifying the gauge connection with the EH spin connection
(standard embedding). As far as the classical invariants of the manifold
are concerned, \ie~Euler characteristic, Hirzebruch signature, indices of the
Dirac and Rarita-Schwinger operators in the gauge singlet representation,
the relevant
computations can be found in the classical literature on the subject [\egh].

New topological invariants are related to the indices of the Dirac operator
in non--singlet representations of the gauge group. We show that the number of
(left--handed) zero--modes of the Dirac operator for
spinors in the fundamental representation
of the instantonic $SU(2)$ gauge group is 1, while for spinors in the adjoint
representation this number is 6. The values taken by these two indices are
respectively related to the existence on this background of 1 self--dual
closed two--form and of 12 deformations of the tangent bundle preserving
the rank and the condition of vanishing of the first Chern class. The
parameters associated with these deformations lend themselves to an
interpretation as collective coordinates of the gauge connection.

The explicit form of all zero--modes is found both with the help of
geometrical intuition and/or by solving directly, when necessary, the Dirac
equation.

\REF\tof{G.~'t~Hooft, Phys. Rev. {\bf D14} (1976) 3422.}
\REF\yaf{A.~Yaffe, Nucl. Phys. {\bf B151} (1979) 247.}
The relevance of the properties of the classical solutions of a field theory
and of the explicit form of the zero--modes of the wave operators
around the corresponding backgrounds needs no further stressing here, if not to
recall that any non--perturbative instantonic saddle--point evaluation of an
Euclidean functional integral would require a detailed knowledge of all these
ingredients [\tof, \yaf].

\REF\rossi{V.~Novikov, M.~Shifman, A.~Vainshtein and
V.~Zakharov, Nucl.Phys. {\bf B260} (1985) 157; \nextline
D.~Amati, G.C.~Rossi and G.~Veneziano, Nucl.Phys. {\bf B249} (1985) 1;\nextline
I.~Affleck, M.~Dine and N.~Seiberg, Nucl.Phys. {\bf B256} (1985) 557;\nextline
D.~Amati, K.~Konishi, Y.~Meurice, G.C.~Rossi and
G.~Veneziano, Phys.Rep. {\bf 162} (1988) 169.}
\REF\wittuno{E.~Witten, Nucl.Phys. {\bf B185} (1981) 513.}
\REF\konuno{K.~Konishi, N.~Magnoli and
H.~Panagopoulos, Nucl.Phys. {\bf B309} (1988) 201; ibid.
{\bf B323} (1989) 441.}
In the case of global supersymmetric theories a detailed study of a possible
breaking mechanism of SUSY via gauge instantons was carried out in [\rossi]
with rather encouraging results. The case of locally supersymmetric theories
was addressed to in [\hp, \wittuno], where it was argued that, in
supergravity, gravitino condensation due to gravitational instantons
can trigger the breaking of SUSY. An important step forward in this line
of arguments was made in [\konuno], where,
in the case of $N=1$ supergravity, the gravitino field strength condensate,
$<\psi_{ab} \psi^{ab}>$, was indeed computed and shown to be finite and
space--time independent. The
calculation was done by performing a saddle--point approximation around the
non trivial classical solution of
the theory represented by the Eguchi--Hanson gravitational instanton [\egh].

The idea behind this kind of approach is the hope of being able to infer
SUSY breaking by putting non trivial condensates in relation with some
anomalous SUSY transformation.
\REF\kontre{K.~Konishi, Phys.Lett. {\bf B135} (1984) 439.}
In global SUSY theories in flat space [\rossi] such a relation exists and
it is called the Konishi anomaly equation [\kontre]. For a Super Yang-Mills
theory coupled to chiral scalar matter
multiplets, $\Phi^i$, belonging to the representations $\underline R_{\>}^i$
of the gauge group, the relevant anomalous commutator reads
$$
\{\bar Q,\bar\lambda_i \phi^j \} = {q^2 c_i\over 32 \pi^2} \chi\chi
\delta_i^j+\phi^j{d \bar W \over d \bar\phi_i}
\eqn\konsei
$$
where $i,j$ are flavour indices, $W$ is the superpotential,
$\bar Q$ is a SUSY charge, $q$ the gauge coupling constant,
$\chi$ is the gaugino, $\phi^i$ and $\lambda^i$ are components of the chiral
multiplet, $\Phi^i$,  and $c_i$ is the index of the representation
$\underline R_i$.
The anomaly equation \konsei~lies in the same supermultiplet as the chiral
anomaly,
\ie~it is a partner of the anomalous divergence equation of the $R$-symmetry
current of the theory. The discovery that (some of) the condensates appearing
in \konsei~acquire a non-zero vacuum expectation value allowed
to derive detailed information on the degeneracy
pattern of the vacuum state manifold and, in some cases (SUSY theories with
no flat directions and chiral matter in suitably choosen representations
of the gauge group) even to conclude that SUSY is
dynamically broken by non--perturbative instanton effects.

To construct a similar argument in supergravity
one has to start with the anomalous divergence of the $R$-symmetry chiral
current, $j^\mu$
$$
D_\mu j^\mu=-{1 \over 384 \pi^2} R_{abcd}\wt R^{abcd}
\eqn\konanomalia
$$
where the tilde stands for the duality operation.\REF\peter{P.K.~Townsend and
P.~van Niewenhuizen, Phys.Rev.{\bf 19D} (1979) 3592.} Since
$R_{abcd}\wt R^{abcd}$ is the top component of a chiral multiplet,
which has $\psi_{ab} \psi^{ab}$ as lowest component [\peter], the analogue of
the anomalous SUSY transformation \konsei~is
$$
\{\bar Q,\bar\lambda \phi\}={ \kappa^2 \over 384\pi^2}\psi_{ab} \psi^{ab}
\eqn\basta
$$
where $\kappa$ is the gravitational coupling constant, $\psi^{ab}$ is the
gravitino field--strength and $\lambda$ and $\phi$ are components of a chiral
matter multiplet (for the sake of definiteness they may be respectively taken
to be
the dilatino and the dilaton). The appearance of the gravitino field--strength
bilinear in the right--hand--side of \basta~led correctly the authors
of [\konuno] to compute the expectation value of $\psi_{ab} \psi^{ab}$,
instead of $\psi_{\m} \psi^{\m}$, as suggested in [\hp].

\REF\taylor{S.-J.~Rey and T.R.~Taylor, Phys.Rev.Lett. {\bf 71} (1993) 1132.}
\REF\noi{M.~Bianchi, F.~Fucito, M.~Martellini and G.C.~Rossi, in preparation.}
Many of the computations we will present in this paper are very much in
the same line of thought of instanton calculus in global supersymmetric
theories [\rossi], where, as we said, the simultaneous presence
of global supersymmetry and of classical instantonic solutions conspire to
give exact space--time independent constant results for certain correlators.
The calculations we present here
are propaedeutical to extend instanton calculus to locally supersymmetric
(\ie~to supergravity) theories [\taylor, \noi].

Contrary to globally supersymmetric
Yang-Mills theories, however, supergravity is not renormalizable.
Strictly speaking, this puts the entire subject of instanton calculus in
supergravity on a rather shaky basis. On this problem we would like to take
the point of view (as also
suggested in [\konuno]) that supergravity theories should be considered
as low energy limits of string theories, which are expected not to suffer
from these deficencies. Thus to the order to
which supergravity theories are formally renormalizable (i.e. generically up to
two loops) results from perturbative and non--perturbative (instanton)
calculations should be considered as the limiting values of the corresponding
exact string results.

There are two more reasons to pursue this philosophy we would like to
mention here. The first has to do with the fact that effective
field theories appear at the moment the only arena where non--perturbative
aspects of string theories can be studied. The second is that,
exploiting the relation between bosonic zero--modes and instanton
``collective coordinates'', explicit computations of the former
may prove to be a useful starting point in the investigation of the structure
of the istanton moduli space over non-compact ALE manifolds.
Except for the purely gravitational sector, this subject is as yet poorly
understood.

The plan of the paper is as follows. In section 2.1 we show that the
effective supergravity theory arising from the heterotic string admits
classical solutions in which the gravitational background is a self--dual
ALE metric and the corresponding gauge background
is constructed by the identifying the gauge connection with the Levi--Civita
spin connection (``standard embedding'').
In the following we will generically refer to these solutions
as ``heterotic ALE instantons''. In section 2.2 we give a world--sheet
interpretation of these classical solutions and we study the relevant Buscher's
duality transformations, showing that heterotic multicenter instantons
are dual to the conformally flat solitonic solutions discovered in [\khuri].
In section 2.3 we specialize the self--dual ALE metric to the Eguchi--Hanson
(EH) gravitational instanton and we collect various
useful formulae for the coupled gauge and gravitational background obtained
in this particular case. In the following we will call this solution the
``heterotic EH instanton''. In section 3 we compute the indices of the gaugino
Dirac operator in the heterotic EH background with the gaugino
in the fundamental and the adjoint representation of the gauge group.
In section 4 we determine the explicit form of the zero--modes
of the operators of the small fluctuations around the heterotic EH background
for all the massless fields of the theory and we discuss the relation
between zero--modes and instanton collective coordinates.
Section 5 contains our conclusions and an outlook of future lines of
investigation. In Appendix A we state our notations and we gather a series of
definitions which should allow the interested reader to reproduce our results.
In Appendix B we report some known facts about the geometry of
four--dimensional ALE manifolds with a special emphasis on the properties
which may be of interest for the study of string propagation on these
backgrounds or for the determination of the structure of their moduli space.

\chapter{Gravitational instantons in heterotic string effective theories}
\section{General Form of Heterotic ALE Istantons}
Our starting point is the $D=10, N=1$ action that arises as a low-energy
effective field theory from the heterotic (or type I) string.
For the massless bosonic fields of the theory (graviton $g_{MN}$,
dilaton $\phi$, antisymmetric tensor $b_{MN}$, vector bosons $A_M$)
the Lagrangian takes the form
$$
L_B = \sqrt g \lbrace - {1\over 2\kappa^2} R(g) -
{1\over 4\kappa^2} (\de\phi)^2 - {e^{-\phi}\over 12 \kappa^2} H^2 -
{ e^{-{\phi\over 2}}\over 4q^2} tr(F^2) \rbrace
\eqn\boselagrangian
$$
The part of the lagrangian quadratic in the fermionic fields (gravitino
$\psi_M$, dilatino $\lambda$, gaugino $\chi$) is given by
$$
\eqalign{
L_{F} &= \sqrt{g} \lbrace -\mezzo \bar\psi_M \Gamma^{MNP} D_N \psi_P
-\mezzo \bar\lambda\Gamma^M D_M\lambda-\mezzo tr(\bar\chi \Gamma^M D_M\chi)\cr
&-{1\over 2\sqrt2} \bar\psi_M \Gamma^N\Gamma^M \lambda \de_N \phi
+ {\sqrt2 e^{-{\phi\over 2}}\over 48}tr(\bar\chi\Gamma^{MNP}\chi)
H_{MNP}\cr
&- {\kappa e^{-{\phi\over 4}}\over 4q}tr(\bar\chi\Gamma^M\Gamma^{NP}F_{NP})
(\psi_M+{\sqrt 2\over 12}\Gamma_M\lambda)
+ {e^{-{\phi\over 2}} \sqrt 2\over 48 }\cr
&(\bar\psi_M\Gamma^{MNPQR}\psi_R+6\bar\psi^N\Gamma^P\psi^Q-\sqrt2\bar\psi_M
\Gamma^{NPQ}\Gamma^M\lambda)H_{NPQ} \rbrace \cr}
\eqn\fermilagrangian
$$
\REF\chaplin{G.F.~Chapline and N.S.~Manton, Phys.Lett. {\bf 120B}
(1983) 105.}
In \boselagrangian~and~\fermilagrangian~$\kappa$ and $q$ are
respectively the gravitational and the gauge coupling constants. $g$
is the determinant of the metric, $R$ is the curvature
scalar and $F$ is the Yang-Mills field--strength.
$H$ is the field--strength of the antisymmetric tensor, modified by the
addition of the Chern--Simons three--form [\chaplin]
$$
\eqalign{
H &= db - {\kappa^2 \over q^2}\Sigma_{YM} \cr
\Sigma_{YM}&=Tr(A\wedge dA+{2\over 3}A\wedge A\wedge A)\cr}
\eqn\torsion
$$
This modification is needed to ensure the invariance of the full action
(including four-fermion terms which for short have not been displayed above)
under local supersymmetry transformations.

It should be noted, however, that in order to
implement the Green-Schwarz  mechanism of anomaly cancellation the standard
way to proceed [\wittdue] is to add to the definition of $H$ the
Chern--Simons three--form for the Lorentz group \foot{By Tr we mean trace
in the adjoint representation of the gauge group. By tr we mean trace
in the vector representation of the Lorentz group}
$$
\Sigma_L = tr(\omega \wedge d\omega +{1\over 3}\omega\wedge \omega\wedge
\omega)
\eqn\sigmal
$$
where $\omega$ is the Levi--Civita spin connection. This redefinition triggers
a chain of further modifications of the Lagrangian and of supersymmetry
transformations which have been partly carried out in [\bdr]. In particular
the authors of [\bdr]~have shown that to first order in $\kappa^2$ it is
necessary to add to the Lagrangian a term quadratic in the Riemann tensor.
Here one has the well known ambiguity related to the possibility of adding
terms proportional to the square of the Ricci tensor or to the
curvature scalar which vanish by the lowest order equations of motion.
Our strategy will be to solve to lowest order in $\ap$ the equations of
motion, but to include the Chern--Simons Lorentz three--form in the
definition of $H$, when we require the validity of the Bianchi identities,
for the sake of showing that certain supersymmetric configurations are
in fact consistent
solutions of the classical equations of motion to all orders in $\ap$.

\REF\fradkin{E.S.~Fradkin and A.A.~Tseytlin, Phys. Lett. {\bf 158B}
(1985) 316.}
In $D=10$, the relation among the inverse string tension $\ap$
(with mass dimension $-2$),
the gravitational constant $\kappa$ (with mass
dimension $-4$) and the gauge coupling $q$ (with mass dimension $-3$)
is $\kappa^2={q^2\ap}$ for the heterotic string (or $\kappa={q^2\over\ap}$ for
the type I superstring). As a result, in \boselagrangian~the $q$-dependence
can be factorized out and reabsorbed into a redefinition of the dilaton
factor, after a
Weyl rescaling of the metric. Eventually the (vacuum expectation
value of the) dilaton factor will play the role of string loop
expansion parameter [\fradkin].

The procedure described above
leads to the so--called $\s$--model variables which in $D$ dimensions are
related to the canonical ones by the equations (capital letters stand
for $\s$--model variables)
$$
\eqalign{
G_{MN} &= e^{{4\phi\over D-2}}g_{MN}, \quad B_{MN} = b_{MN},\quad A_M = A_M\cr
\Psi_{M} &= e^{{\phi\over D-2}}(\psi_M-{\Gamma_M\over 2\sqrt2}\lambda)
\quad \Lambda = e^{{\phi\over D-2}} \lambda
\quad X = e^{{\phi\over D-2}}\chi \cr}
\eqn\variables
$$
In terms of the $\s$--model variables the bosonic part of the lagrangian
becomes [\cfmp]
$$
{\cal L}_B = {\sqrt{G} \over 2\ap} e^{-2\phi}
(R - 4(\de \phi)^2 + {1\over 12} H^2 + {\ap \over 2} F^2)
\eqn\sigmalagrangian
$$
where $G$ is the determinant of the metric $G_{MN}$.
The Euler-Lagrange equations obtained by varying \sigmalagrangian~are
$$
\eqalign{
R_{MN} - \mezzo G_{MN} R &= \ap T_{MN} \cr
D_M (e^{-2\phi} F^{MN})&=A_M D_P(e^{-2\phi} H^{MNP}) \cr
D_M D^M\phi + {1\over 6} e^{-2\phi} H^2 &= \ap \{ F(A)^2 -
R(\Omega_{-})^2\} \cr
D_M(e^{-2\phi} H^{MNP}) &= 0 }
\eqn\motouno
$$
where $T_{MN}$ is the total stress energy tensor and $D_M$ the fully
covariant derivative. In presence of a non-zero torsion
one can define the two generalized spin connections
$$
\Omega_{M\pm}{}^{AB}=\omega_{M}{}^{AB}\pm H_M{}^{AB}
\eqn\omegapm
$$
where $A, B$ are frame indices. Note that it is the generalized spin connection
$\Omega_{-}$, which enters in \motouno.

We look for a supersymmetric solution of the above equations whose
instantaneous time slice is five--dimensional [\penta].
We split the $10$ dimensional coordinates $\{x^M\}$ in two sets:
a six--dimensional one
$\{y^i; i=0,1,\cdots,5\}$ and a four--dimensional Euclidean part
$\{x^\m; \m=0,1,2,3\}$\rlap.\foot{In spite of the fact that throughout
this paper we
will stick to four--dimensional metrics with Euclidean signature, we will label
space--time coordinates with the index $\mu$ running from $0$ to $3$}
We assume that the fields do
not depend on the $y$-coordinates and that the only non--trivial components
of all tensors are those spanning the four--dimensional transverse space.
With this restriction the sum of the dilaton, axion and vector boson euclidean
stress energy tensor becomes
$$
\eqalign{
T_{\mn} &=
{1\over 4} G^{\r\s} (F_{\m\r}+ \wt F_{\m\r}) (F_{\n\s}- \wt F_{\n\s})
-{1\over 8} G_{\mn} ((\de\phi)^2 -  (\wt H)^2)\cr
&+ {1\over 2 \ap} (\de_\m \phi +  \wt H_\m)
(\de_\n \phi - \wt H_\n)+ {1\over 2 \ap} (\de_\m \phi -  \wt H_\m)
(\de_\n \phi + \wt H_\n) \cr}
\eqn\stresstensor
$$
where the tilde indicates the dual of the field--strengths $F$ and $H$.

An ansatz for the solution of the bosonic equations of motion, for which the
stress energy tensor \stresstensor~vanishes, is given by [\calluno, \khuri,
\penta]
$$
\eqalign{
F_{\mu\nu} &= \pm \wt F_{\mn} =\pm {1 \over 2} \sqrt{G}
\varepsilon_{\mn}{}^{\rho\sigma}F_{\rho\sigma} \cr
G_{\mn} &=  e^{2\phi} \wh g_{\mn} \cr
H_{\mnr}&= \pm \sqrt{G} \varepsilon_{\mnr}{}^\sigma\de_\sigma\phi \cr}
\eqn\forsol
$$
with $\wh g_{\mn}$ any arbitrary Ricci flat metric. The fermionic equations
of motion are easily satisfied by setting all fermion fields equal to zero
[\wittdue,
\candelas].

In order to find under which conditions this solution is supersymmetric,
we observe that the local SUSY variations of the bosonic fields are all
trivially zero, due to the above vanishing of the background fermion fields,
while for the gravitino, dilatino and gaugino field variations one has
$$
\eqalign{
\delta\psi_\mu &= (\de_\mu-{1 \over 4} \Omega_{\mu+}{}^{ab}
\gamma_{ab})\epsilon + \ap\delta_2\psi_\mu \cr
\delta\lambda &= -{1 \over 4}(\gamma^{\mu}\de_\mu \phi -{1 \over 6}
H_{\mu\nu\rho} \gamma^{\mu\nu\rho})\epsilon+\alpha^\prime\delta_2\lambda \cr
\delta\chi &= -{1 \over 4}F_{\mu\nu}\gamma^{\mu\nu}\epsilon+
\alpha^\prime\delta_2\chi \cr }
\eqn\varuno
$$
where $\gamma_{\mn}\ldots$ are the usual antisymmetric products of
gamma matrices (see Appendix A). Greek and latin letters denote
four--dimensional coordinates and frame indices respectively.
In the background we are considering the parts of the variations indicated by
$\delta_2$, being  quadratic in the fermi fields, are all zero. It is also
easy to see
that, if we take the plus (minus) sign in \forsol~and a right-(left-)handed
spinor, $\epsilon$, the variation of the gaugino is zero, as $F_{\mn}$ is
(anti)self--dual. Similarly the dilatino variation is zero, thanks to the
ansatz for $H_{\mnr}$. The generalized spin connections (with torsion) of
the metric $G_{\mn}$ become
$$
\Omega_{\mu\pm}{}^{ab} = \wh\omega_\mu{}^{ab}-(\wh e_\mu{}^a \wh e^{b \nu}
-\wh e^{a \nu} \wh e_\mu^b)\de_\nu \phi \pm H_\mu{}^{ab}
\equiv\wh\omega_\mu{}^{ab}+w_\mu{}^{ab} \pm H_\mu{}^{ab}
\eqn\spinconn
$$
where hatted quantities are computed with respect to the metric $\wh g_{\mn}$.
Using the duality relation ${1 \over 2}
\varepsilon^{ab}{}_{cd}w_\mu{}^{cd}= H_\mu{}^{ab}$ (${1 \over 2}
\varepsilon^{ab}{}_{cd}w_\mu{}^{cd}= - H_\mu{}^{ab}$),
which again is a consequence
of the ansatz \forsol, we see that the gravitino variation may be
set to zero if $\wh\omega_\mu{}^{ab}$ is taken to be (anti)self--dual.
With this choice $\Omega_{\m+}{}^{ab}$ is (anti)self--dual, while
$\Omega_{\m-}{}^{ab}$ has in general no definite duality.
{}From now on we will specialize to the self--dual case by choosing the + sign
in \forsol~and $\epsilon$ right--handed in \varuno.

To have a consistent supersymmetric solution,
we must require the expressions \forsol~for $F$ and $H$
to be the field--strengths of the fundamental bosonic fields,
$A_{\m}$ and $B_{\m\n}$ respectively. This amounts to impose the
validity of the (modified) Bianchi identities
$$ \eqalign{
 dF &+A\wedge F =0 \cr
 dH &=\ap \{ trR(\Omega_{-})\wedge R(\Omega_{-})-
{1 \over 30} TrF(A)\wedge F(A) \} \cr}
\eqn\bianchi
$$
where, as we said, in $dH$ we also have included the contribution of the
Chern--Simons Lorentz three--form.
To satisfy \bianchi~it is natural to relate the gauge and spin connection
through a generalized  form of ``standard embedding''.

Given the existence of a self--dual and an antiself--dual 't Hooft symbol,
$\eta^i_{ab}$ and $\bar\eta^i_{ab}$, and of two generalized spin connections,
$\Omega_{\pm}$, four gauge connections could in principle be constructed,
namely
$$ \eqalignno{
 A^i{}_\mu &={1 \over 2} \eta^i{}_{ab} \Omega_{\m-}{}^{ab}
= {1 \over 2} \eta^i{}_{ab} \wh\omega_{\m}{}^{ab}
& \eqnalign{\embuno}\cr
A^i{}_\mu &={1 \over 2} \bar\eta^i{}_{ab} \Omega_{\m-}{}^{ab}
= {1 \over 2} \bar\eta^i{}_{ab} (w_{\m}{}^{ab} - H_{\m}{}^{ab})
& \eqnalign{\embdue}\cr
A^i{}_\mu &={1 \over 2} \eta^i{}_{ab} \Omega_{\m+}{}^{ab}
& \eqnalign{\embtre}\cr
A^i{}_\mu &={1 \over 2} \bar\eta^i{}_{ab} \Omega_{\m+}{}^{ab} =0
& \eqnalign{\embqua}\cr}
$$
with $i=1,2,3$ the index of an $SU(2)$ subgroup of the heterotic gauge
group.

Of the above four cases, the last one, \embqua, is trivial, because
$\bar\eta^i{}_{ab}$
and $\Omega_{+}$ have opposite duality. Equation \embtre~does not lead to
curvatures with
definite duality, because the self--duality of $\Omega_{+}$ guarantees the
self--duality of the Riemann tensor $R_{\mn}{}^{ab}(\Omega_{+})$ with respect
to frame indices, but not necessarily with respect to coordinate indices.
This is easily seen by making use of the modified exchange
property of the Riemann curvature tensor
$$
 R_{\mn}{}^{ab}(\Omega_{+}) = R^{ab}{}_{\mn}(\Omega_{-}) =
 e^{a\r} e^{b\s} e_{c\m} e_{d\n} R_{\r\s}{}^{cd}(\Omega_{-})
\eqn\flip
$$
valid when the torsion tensor is closed, \ie~when $dH=0$.
Since $\Omega_-$ and, hence, $R(\Omega_-)$ have no definite duality,
the gauge field--strength resulting from the use of \embtre
$$
F^i{}_{\mn} = {1 \over 2}\eta^i{}_{ab} R_{\mn}{}^{ab}(\Omega_+)
\eqn\fieldstrength
$$
is not self--dual.
As for the other two possibilities, \embuno~and \embdue, they are both good
starting points for a perturbative (with respect to $\ap$) solution of the
equations of motion.
To enforce the vanishing of $\ap$ corrections (at least up to second and third
order) the condition $dH=0$ must be enforced. This requires the validity of the
relation
$$
trR(\Omega_{-})\wedge R(\Omega_{-})={1 \over 30} TrF(A)\wedge F(A)
\eqn\cancellation
$$
which is not satisfied in general, unless one further restricts
either the form of the metric $\wh g_{\mn}$ and/or the expression of the
dilaton.

The first possibility has been considered in [\calluno] and [\khuri] with the
conclusions that the non conformal part of the metric must be flat
(in our notations this means $\wh g_{\mn}=\delta_{\mn}$)
and that, given the form of the supersymmetric
ansatz \forsol~for $H$, the Bianchi identity, $dH=0$, requires the
dilaton factor to obey the Laplace-Beltrami equation in flat background, \ie
$$
e^{-2 \phi}\de^\mu \de_\mu e^{2 \phi} =0
\eqn\laplace
$$

In this paper we would like to concentrate on the second option. The
choice \embuno~is consistent with the identification of $F$
with $R(\Omega_{-})$ implied by \cancellation, only if the dilaton is
constant and the torsion is zero. In this case, the solution \forsol~is
completely specified by the choice of a self--dual metric $\wh g_{\mn}$.

In sect.~2.2 we will prove that the class of
solutions found in [\khuri] (corresponding to the choice \embdue)
are ``dual'' in the sense of [\buscher] to the class of solutions of
our interest here.

\REF\gh{ {\it Euclidean Quantum Gravity}, eds. G.~Gibbons and
S.W.~Hawking, World Scientific Co., 1993.}
To make more explicit the ansatz \forsol~we will specify the form of
the self--dual background metric, $\wh g_{\mn}$.
Several classes of self--dual metrics are known in literature [\egh, \gh].
\REF\ghp{G.W.~Gibbons, S.W.~Hawking and M.J.~Perry, Nucl. Phys. {\bf B138}
(1978) 141;\nextline
G.W.~Gibbons and C.N.~Pope, Comm. Math. Phys. {\bf 66} (1979) 267.}
We will concentrate on Asymptotically
Locally Euclidean metrics. Our main motivation is that for generic
(even not self--dual) ALE metrics the
euclidean  Einstein-Hilbert action is conjectured to be
semipositive definite [\ghp]. This is of course a necessary prerequisite for
the sake of giving a mathematically sound definition of a quantum
theory of gravity in terms of functional integrals.

Geometrical properties of four--dimensional ALE manifolds are well known
in literature [\egh] and are recalled for completeness in Appendix B.

\section{World--sheet Interpretation of heterotic ALE instantons}
In this section we would like to give a world--sheet interpretation to the
propagation of the string in the heterotic ALE instanton backgrounds
we have previously described, that is to say, we want to study the
corresponding $\s$--models which describe the propagation of the string
degrees of freedom
associated to the four--dimensional Euclidean $x$--space.

Apart from the bosonic coordinates of the string, denoted by $X^\m$,
one has to consider their superpartners, $\Psi^\m$, which from the
two--dimensional point of view are left--handed fermions, and, in order to
include the extra heterotic degrees of freedom, four additional right--handed
fermions, which we denote by $\bar \Lambda^a$.

The action of the two--dimensional $\s$--model describing the propagation
of the
heterotic string in a background with metric $G_{\mn}$, antisymmetric tensor
$B_{\mn}$ and gauge fields $A^i{}_\m$ is given by [\cfmp, \caltho]
$$
\eqalign{
&S = {1\over 4\pi \ap}\int dz d {\bar z} \lbrace
G_{\mn}(X)\de X^\m\bar\de X^\n + B_{\mn}(X)\de X^\m\bar\de X^\n \cr
&+ iG_{\mn}(X)\Psi^\m\bar D_{-}\Psi^\n
+ i\delta_{ab}\bar\Lambda^a D_{+}\bar\Lambda^b
+ F^i_{\mn}(A)\eta_{iab}\Psi^\m\Psi^\n\bar\Lambda^a\bar\Lambda^b
\rbrace \cr}
\eqn\sigmamodel
$$
where the covariant derivatives are defined by ($\de = \de_z$,
${\bar \de} = \de_{\bar z}$)
$$
\eqalign{
\bar D_{-}\Psi^a &= \bar\de\Psi^a + \Omega_{\n-}^{ab}\bar\de X^\n\Psi_b\cr
D_{+}\bar\Lambda^b &= \de\Lambda^a + A^i{}_{\n}\eta_i{}^{ab}
\de X^\n\Lambda_b\cr}
\eqn\derivative
$$
with $\Omega_{\n-}^{ab} = \wh\omega_{\n}{}^{ab}-H_{\n}{}^{ab}$ the spin
connection with torsion. In \sigmamodel~we have neglected the dilaton term
[\fradkin], because it is of higher order in $\ap$.

\REF\hklr{N.J.~Hitchin, A.~Karlhede, U.~Lindstr\"om and
M.~Ro\v cek, Comm.Math.Phys. {\bf 108} (1987) 535.}
When the metric is hyperk\"ahlerian [\hklr, \sigmaquattro],
the $\s$--model \sigmamodel~is invariant under extended supersymmetries.
In particular this is true, if one takes $G_{\mn}$ to be an ALE metric and
sets $B_{\mn}$ to zero, as in the solution of our interest here.
Besides the usual left--handed supersymmetry which on the bosonic
coordinates $X^\m$ induces the transformation
$$
\delta X^\m = \epsilon^{(0)} \Psi^\m
\eqn\susyvarbose
$$
\sigmamodel~is invariant under three extra supersymmetries [\sigmaquattro]
whose action on the $X^\m$ is
$$
\delta X^\m = \epsilon^{(i)} J_i{}^\m{}_\n \Psi^\n \quad i=1,2,3
\eqn\susyextra
$$
with the $J$'s the three covariantly constant complex structures defined
in (B.4). All of this is in agreement with the known result
that non-linear $\s$--models on hyperk\"ahler manifolds (a class of manifolds
which includes four--dimensional ALE manifolds, see Appendix B) admit
four left--handed supersymmetries. Actually, thanks to \embuno,
\sigmamodel~admits four right--handed extra supersymmetries which can be
exposed by interpreting the fermions $\bar\Lambda^a$ as (frame components of)
right--handed superpartners of the $X^\m$. Indeed it can be shown
that \sigmamodel~is invariant also under the four additional
supersymmetric transformations whose action on the $X^\m$ is
$$
\delta X^\m = \bar\epsilon^{(A)}\bar J_A{}^\m{}_\n E^\n{}_a \bar \Lambda^a
\quad A=0,1,2,3
\eqn\susyright
$$
where the $E^\n{}_a$ are the components of the $\s$--model tetrad and
$\bar J_0{}^\m{}_\n = \delta^\m{}_\n$.

This proves that \sigmamodel~with the standard embedding \embuno~and
constant dilaton has $N=(4,4)$ supersymmetries on the world--sheet.
An analogous observation was made in [\calluno, \khuri] with the
conclusion that lowest order solutions of the string equations of
motion, built from the standard embedding \embdue, are exact
solutions to all orders in $\ap$, because of
the expected absence of radiative corrections in the corresponding
two--dimensional $N=(4,4)$ $\s$--models.

In fact the authors of [\calluno] could even go further and exhibit,
in the limiting case of an asymptotically vanishing dilaton,
an exact conformal field theory
which describes the string propagation on a semi-wormhole
geometry coupled to a 't Hooft instanton with winding number $\n=1$.
{}From now on we will call this solution the CHS solution.

The main difference between the heterotic ALE solutions we have described
in sect.~2.1 and the CHS solution
lies in the non triviality of the metric. Since the conformal factor of the
$\s$--model metric of the CHS solution is the exponential of the dilaton field,
the corresponding ``canonical'' metric is flat\rlap.\foot{For product spaces,
like $(M_4\otimes M_6)$, with block diagonal metrics,
as the ones we are considering
here, the appropriate Weyl rescalings are given by \variables~with $D=4$}
In this case,
even in the presence of a non vanishing axionic torsion, as it
happens with the ansatz \forsol, no
gravitino zero--modes are expected which could give rise to a non trivial
gravitino condensate.

In the sense of Buscher's duality [\buscher] ALE instantons or, more in
general, self--dual multi--center metrics are tightly
related to the multi--monopole solutions of [\khuri].
Starting from the fact that the dilaton of the multi-monopole
solution has exactly the form of
the potential $V(\vec x)$ in the Gibbons--Hawking metric (B.1) (see (B.3)),
one can go from the self--dual multi--center
metrics to the conformally flat multi--monopole solutions.

\REF\alvarez{E.~Alvarez, L.~Alvarez-Gaum\'e and Y.~Lozano,
{\it On Non-Abelian Duality}, CERN-TH-7204/94.}
Duality with respect to both abelian and
non-abelian isometries for self--dual metrics have already been
considered [\alvarez].
Without going into details and referring directly to
[\buscher, \alvarez], one can show that a $\s$--model such as \sigmamodel~can
be related to a dual $\s$--model by exploiting the existence
of isometries of the initial metric. In the case of the
multi--center metric (B.1) the relevant $U(1)$ isometry is generated by
the Killing vector ${\de \over \de \tau}$.
The formulae for the transformation to the dual metric, dilaton and
antisymmetric tensor fields are given by
$$
\eqalign{
G^{\du}_{00} &= {1\over G_{00}},\quad
G^{\du}_{0i} =  {B_{0i}\over G_{00}},\quad
G^{\du}_{ij} =  G_{ij}-{1\over G_{00}}(G_{0i}B_{0j}+G_{0j}B_{0i})\cr
B^{\du}_{0i} &=  {G_{0i}\over G_{00}},\quad
B^{\du}_{ij} =  B_{ij}+{1\over G_{00}}(G_{0i}B_{0j}-G_{0j}B_{0i})\cr
\Phi^{\du} &= \Phi - \mezzo \log(G_{00})\cr}
\eqn\duality
$$
where D stands for dual. The index $0$ refers to the cyclic coordinate
$\tau$, while the
index $i$ is relative to the other three cartesian coordinates, $x^i$.
Using the explicit expression of the metric (B.1), which is
already expressed in adapted coordinates, one gets
$$
\eqalign{
G^{\du}_{00} &= V(\vec x),\quad G^{\du}_{0i} = 0,\quad G^{\du}_{ij}
= V(\vec x) \delta_{ij}\cr
B^{\du}_{0i} &= \omega_i,\quad B^{\du}_{ij} = 0\cr
\Phi^{\du} &= \Phi_0 + \mezzo \log(V(\vec x))\cr}
\eqn\explicitdual
$$
The gauge field $A^{\du}_\m$ is still obtained by using the standard embedding
equation, whose validity is preserved by the duality transformation.
As it is evident from \explicitdual, the dual metric is conformally flat with
a conformal factor which
exactly equals the exponential of the dilaton.
Thus, as announced, the canonical metric turns out to be flat,
all the non triviality of the solution being tranferred to the dilaton-axion
system. It is nice to see that, thanks to the standard embedding and the
peculiar form of multi--center metric following from (B.2), the dilaton-axion
system satisfies the condition for a supersymmetric background, \ie
$$
H^{\du}_{\mnr}= \sqrt{G^{\du}} \varepsilon_{\mnr}{}^\s\de_\s\Phi^{\du}
$$
and that the resulting gauge field--strength $F^{\du}_{\mn}$ is self--dual.

For $\epsilon=1$ the dual backgrounds coincide with the multi-monopole
solutions found in [\khuri], which in this way appear to be dual to
Euclidean multi-Taub-NUT spaces. Connections on Euclidean multi-Taub-NUT
spaces are self--dual and admit a ``standard embedding'',
similarly to ALE spaces [\char]. Notice, however, that multi-Taub-NUT spaces
are asymptotically Euclidean only in three directions.

Even if the solutions we have discussed above are conjectured to
be exact solutions to all orders in $\ap$, it seems very difficult
to find an exact ``algebraic'' conformal field theory which would describe
the string propagation on these self--dual backgrounds (and their dual
theories) at generic point of their moduli space.
Further duality transformations with respect to non-abelian isometries
[\alvarez] do not seem to give any clue for the solution of the problem.

\REF\torinesi{D.~Anselmi,
M.~Billo', P.~Fre', L.~Girardello and A.~Zaffaroni, {\it ALE Manifolds and
Conformal Field Theories}, SISSA/44/92/EP, hep-th/9304135}
In spite of these difficulties, $\s$--models with ALE
instanton backgrounds in a particular singular limit,
have been identified with solvable $\complex^2/\Gamma$ orbifold conformal
field theories, $\Gamma$ being a Kleinian subgroup of $SU(2)$ [\torinesi].
In order to recover the propagation of the string on a smooth manifold
one must perturb the orbifold conformal field theory along the ``truly
marginal deformations'' given by the moduli of the (orbifold) background. The
authors of [\torinesi] have shown that there exists a precise correspondence
between the ``short representations'' of the $N=4$ superconformal algebra and
the moduli of metric and antisymmetric tensor, although they have
not been able to determine the exact transformation connecting ``geometric''
and ``$\s$--model'' bases. Notice that models perturbed along marginal
directions correspond to superconformal field theories with $c=(6,6)$
(which is the value of the conformal anomaly of four flat supercoordinates).

In the construction of a solution to the classical equations of motion
of the $D=10$ heterotic (or type I) superstring theory, we have been,
so far, rather cavalier about the dynamics and the geometry of the extra
six dimensions, $\{y^i; i=0,1,\cdots,5\}$.
On this question we essentially have
two alternatives. One is to interpret our four--dimensional ALE solutions
in terms of the propagation of a flat five-brane in a five--dimensional
space--time, $y_0$ being the time coordinate. In this picture the complete
ten--dimensional solution would correspond to an $N=(4,4)$ $\s$--model,
whose action is the sum of a non trivial part (eq.~\sigmamodel) plus
a free part describing the dynamics of the six--dimensional $y$--coordinates.
In this case the four--dimensional effective field theory will have
$N=2$ space--time supersymmetry.

Alternatively, in view of a possible application to the study of dynamical SUSY
breaking, it seems more appealing to interpret our heterotic ALE
instantons as solutions
of the Euclidean equations of motion of a $D=4,\> N=1$ effective
supersymmetric theory.
In this case the six--dimensional space is taken to be a
Calabi-Yau threefold. The corresponding internal conformal field theory
has $N=(2,2)$ supersymmetry. The spectrum of massless
fields is consequently enlarged, but if we embed the istantonic
$SU(2)$ group in the hidden $E(8)$, the ALE solution would still remain a
solution of the equations of motion. In fact the ALE solution only involves
space--time fields arising from the universal (``identity'') sector of
the internal superconformal field theory which is independent of the
details of the compactification.

\section{A Particular Solution: the Heterotic Eguchi--Hanson instanton}
\REF\prasad{M.K.~Prasad, Phys.Lett. {\bf 83B} (1979) 310.}
In the previous subsection we have found a general class of solutions of the
effective low energy theory arising from the underlying heterotic
string theory. We now specialize our solution to the
Eguchi--Hanson gravitational instanton [\eh] which corresponds
to a two center metric ($k=1$) with $\epsilon=0$ and $m=\mezzo$
in (B.3). Placing the two centers, $\vec x_1, \vec x_2$, symmetrically
along the $z$ axis, at a distance $|\vec x_1 -\vec x_2|={a^2 \over 4}$,
and performing the change of variables $(\vec x\equiv(x,y,z))$ [\prasad]
$$
\eqalign{
x &= {u^2\over 8} \sin{\theta} \cos{\psi} \quad\quad
z ={r^2 \over 8} \cos{\theta} \cr
y &= {u^2\over 8} \sin{\theta} \sin{\psi} \quad\quad
\tau = 2\varphi \cr}
\eqn\trasprasad
$$
the line element (B.1) can be cast in the form
$$
ds^2= \wh g_{\mn} dx^\mu dx^\nu=
{dr^2 \over \unoar}+r^2 (\sigma_x^2+\sigma^2_y+(\unoar)\sigma_z^2)
\eqn\ehmetric
$$
where the radial variable is constrained
by $r\geq a$ and we have introduced the definition
$u=r\sqrt{\unoar}, 0\leq u \leq \infty$.
The $\sigma$'s are the left-invariant one--forms of the $SU(2)$ group
and are defined in Appendix A. The global topology of the EH manifold,
$M_{EH}$, is that of $T^*(S^2)$ (the cotangent bundle over the Riemann
two-sphere). The slices at fixed $r$ have the topology of a
distorted sphere (at $r \approx a M_{EH}\approx \real^2 \times S^2$,
at $r \approx \infty M_{EH}\approx \real \times S^3/Z_2$)
with points, antipodal with respect to the origin,
identified. This identification is necessary to remove the (apparent) ``bolt"
singularity at $r=a$. The boundary at $\infty$ is a
real projective space, \ie~$RP_3=S^3/Z_2$.

The group of isometries of $M_{EH}$ is the subgroup $SU(2)_R\otimes U(1)_L$
of the Euclidean Lorentz group,
$O(4)\approx SU(2)_R\otimes SU(2)_L$. The
generators of $SU(2)_R\otimes U(1)_L$ are
given in terms of the four linearly independent Killing vectors, $\bar
\xi_{(i)}^{\m}, i=1,2,3$ and $\xi_{(3)}^{\m}$ (see (A.15), (A.19)),
by the formulae
$$
\eqalign{
\bar \xi_{(1)}^{\m} \de_\m &=  (\sin{\varphi} {\ddtheta} -
     {\cos{\varphi}\over \sin{\theta}} {\ddpsi}
+ {\cos{\theta} \cos{\varphi} \over \sin{\theta}} {\ddphi}),\cr
\bar \xi_{(2)}^{\m} \de_\m &=  - ( \cos{\varphi} {\ddtheta} +
     {\sin{\varphi}\over \sin{\theta}} {\ddpsi}
- {\cos{\theta} \sin{\varphi} \over \sin{\theta}} {\ddphi}),\cr
\bar \xi_{(3)}^{\m} \de_\m &=  {\ddphi}, \cr
\xi_{(3)}^{\m} \de_\m &=   {\ddpsi} \cr}
\eqn\killinguno
$$
where $0 \leq \theta \leq \pi$, $0 \leq \phi \leq 2\pi$,
$0 \leq \psi \leq 2\pi$. The restricted range of $\psi$ is related to the
$Z_2$ identification of antipodal points mentioned above.
Indeed referring to the relation (A.13) between cartesian and spherical
coordinates, it is immediately seen that the identification of points
having $\psi$ differing by (multiples of) $2\pi$ implies the identification of
the points of the manifold that have opposite coordinates.

For completeness we also report the definition of the generators of the coset
space $SU(2)_L /U(1)_L$
$$
\eqalign{
\xi_{(1)}^{\m} \de_\m &= (\sin{\psi} {\ddtheta} -
{\cos{\psi}\over \sin{\theta}} {\ddphi}
+ {\cos{\theta} \cos{\psi} \over \sin{\theta}} {\ddpsi})\cr
\xi_{(2)}^{\m} \de_\m &= - ( \cos{\psi} {\ddtheta}
+ {\sin{\psi}\over \sin{\theta}} {\ddphi}
- {\cos{\theta} \sin{\psi} \over \sin{\theta}} {\ddpsi})\cr}
\eqn\killingdue
$$
These operators correspond to global symmetries of the classical lagrangian,
broken by the EH background.

The EH tetrad one--forms are
$$
e^a\equiv e^a_\mu dx^\mu=\{ {r \over u} dr, r \sigma_x, r \sigma_y,
u \sigma_z \}
\eqn\tetrad
$$
The self--dual spin connection one--forms,
$\omega^{ab}\equiv\omega^{ab}_\mu dx^\mu
= \mezzo \epsilon^{ab}{}_{cd} \omega^{cd}$,
are related, thanks to \embuno, to the gauge connection one--forms,
$A^{i}\equiv A^{i}_\mu dx^\mu$ ($i=1,2,3$ is the index of the $SU(2)$ adjoint
representation). Using \ehmetric, one explicitely finds
$$
\eqalign{
\omega^{10} =\omega^{23}&={1 \over 2} A^1 = {u \over r}\sigma_x \cr
\omega^{20} =\omega^{31}&={1 \over 2} A^2 = {u \over r}\sigma_y \cr
\omega^{30} =\omega^{12}&={1 \over 2} A^3 = {u u' \over r}\sigma_z \cr}
\eqn\spinconndue
$$
where the prime means derivative with respect to $r$.
For later use, we notice the identity
$$
{i\over 2} A^k_\m \s_k = {1\over 4} \omega^{ab}_\m \s_{ab}
\eqn\connmatrix
$$
which immediately follows from \embuno~and (A.4), \ie~from the fact that
through \embuno~we are identifying the gauge connection, $A^i$, which belongs
to the adjoint representation of the instantonic $SU(2)$ with the spin
connection, $\omega^{ab}$, which similarly belongs to the adjoint
representation of the $SU(2)_L$ subgroup of the Euclidean Lorentz group.

We furthermore introduce the coefficient functions, $a^{(k)}$, defined by
$$
A^k_\m = a^{(k)} e^k_\m \quad k=1,2,3
\eqn\coefficient
$$
{}From \tetrad~and \spinconndue~one explicitly gets
$$
a^{(1)} = a^{(2)} = {2u\over r^2} \quad a^{(3)} = {2u'\over r}
\eqn\explicit
$$
Finally we record the expressions of the curvature two--forms,
$R^{ab} \equiv \mezzo R^{ab}_{\m\n} dx^\m dx^\n$ and
$F^i \equiv \mezzo F^i_{\m\n} dx^\m dx^\n$
$$
\eqalign{
R^{10}=R^{23}&={1\over 2}F^1={2\over r^2}\ar(e^0\wedge e^1-e^2\wedge e^3)\cr
R^{20}=R^{31}&={1\over 2}F^2={2\over r^2}\ar(e^0\wedge e^2-e^3\wedge e^1)\cr
R^{30}=R^{12}&={1\over 2}F^3=-{4\over r^2}\ar(e^0\wedge e^3-e^1\wedge e^2)
\cr}
\eqn\curvat
$$

\chapter{Index Theorems for the Heterotic Eguchi--Hanson Instanton}
In this section we want to study the topological properties of
the particular ALE solution, the heterotic EH instanton, described
in sect.~2.3, and compute the indices of all
relevant wave operators, using standard formulae from index theorems.

\REF\duffdue{A.J.~Hanson and H.~R\"omer, Phys.Lett.{\bf 80B} (1978) 58;
\nextline S.M.~Christensen and M.J.~Duff, Nucl.Phys. {\bf B154} (1979) 301;
\nextline H.~R\"omer, Phys.Lett.{\bf 83B} (1979) 172. }
For the gravitational part all topological invariants have been computed
in [\egh, \duffdue].
The Euler characteristic, $\chi$, has both a ``bulk'' and a ``boundary''
contribution and it is given by the formula
$$
\chi = {1 \over 32 \pi^2} \int_M \varepsilon^{abcd}( R_{ab}\wedge R_{cd})
-{1 \over 16 \pi^2} \int_{\de M} \varepsilon^{abcd}( \theta_{ab}\wedge R_{cd}
 - {2\over 3} \theta_{ab}\wedge\theta_{ce}\wedge\theta^e _d)
\eqn\euler
$$
with $\theta_{ab}$ the second fundamental forms of the Lorentz group
\ie~the difference between the spin connection of the original metric,
computed on the boundary, and the spin connection
obtained from the boundary product-metric.

If we take $M=M_{EH}$\foot{From now
on throughout this section the subscript $EH$ on $M$ will always be
understood},
the boundary, $\de M$, will be represented by a slice at $r=r_0$, which in
the end we will have to send to $\infty$. Since the EH metric
factorizes on $\de M$ one easily gets
$$
\eqalign{
\theta_{01}&=-(1-({a \over r_0})^4)^{\mezzo}\s_x \cr
\theta_{02}&=-(1-({a \over r_0})^4)^{\mezzo}\s_y \cr
\theta_{03}&=-(1+({a \over r_0})^4)\s_z \cr
\theta_{12}&=\theta_{23}=\theta_{31}=0 \cr}
\eqn\seconform
$$
It follows from \euler~and \seconform
$$
\chi_{\scriptscriptstyle{EH}}={3\over 2}+\mezzo=2
\eqn\chiEH
$$

The Hirzebruch signature, $\tau$, receives contribution only from
the ``bulk'' and it is given by
$$
\tau = -{1 \over 24 \pi^2} \int_M tr( R\wedge R)=b_2^+-b_2^-
\eqn\hirz
$$
$\t$ represents the difference between the number of normalizable
antiself--dual and self--dual harmonic two--forms in the manifold.
For the EH case one gets
$$
\t_{\scriptscriptstyle{EH}} = -1
\eqn\tauEH
$$
The number of (left--handed) gravitino zero--modes, $\psi_\m^\a$,
\ie~the number of solutions of the equation
$$
\epsilon^{\l\mnr}\bar\s_\m D_\n \psi_\r =
\epsilon^{\l\mnr}\bar\s_\m(\de_\n+{1 \over 4}\omega_\n{}^{ab}
\sigma_{ab})\psi_\r=0
\eqn\raritaschwinger
$$
turns out to be $2|\tau|= 2$ in the EH case. The explicit
expression of the two gravitino zero--modes will be presented in sect.~4.

The Dirac equation for the gauge-singlet (left-handed) dilatino field, $\l^\a$,
$$
\fey_S \lambda \equiv \bar \s^\m D_\m \lambda \equiv
\bar\s^\mu (\de_\mu + {1 \over 4}\omega_\m{}^{ab} \sigma_{ab})\lambda =0
\eqn\dilatino
$$
has no normalizable solutions. This result immediately follows from the
observation that in the Ricci-flat EH background
$$
\fey_S\fey_S|_{EH} = (D^2 + {1 \over 8} R_{\mn a b} \s^{\mn} \s^{ab})|_{EH}
= D^2|_{EH}
$$
if one remembers that in a non-compact
background the equation $D^2\phi=0$ has no normalizable solutions.

Finally the expression of the Dirac operator acting on the
(left--handed) gaugino, $\chi^\a$, is
$$
\fey \chi^I = \bar\s^\mu\{(\de_\mu + {1 \over 4}\omega_\m{}^{ab}
\sigma_{ab})
\delta^I{}_K+{i \over 2} f^I{}_{JK}
A^J{}_\mu \}\chi^K=0
\eqn\gaugino
$$
where $f_{IJK}$ are the structure constants of the gauge group.

Since the istanton lives in an $SU(2)$ subgroup of the heterotic
$E(8)\otimes E(8)$ (or $SO(32)$) group,
it is necessary to decompose its adjoint representation
in terms of representations of the istantonic $SU(2)$. As we have argued
in sect.~2.2, it is convenient to embed the instantonic $SU(2)$ in
the ``hidden'' $E(8)$. By decomposing this $E(8)$ with respect to its maximal
subgroup, $SU(2)\otimes E(7)$, the 248--dimensional adjoint representation
of $E(8)$ breaks as
$$
\underline{248}=(\underline3,\underline1)\oplus
(\underline2,\underline{56})\oplus (\underline1,\underline{133})
\eqn\decomposition
$$
We see that, for the maximal embedding we are considering here,
only the fundamental and the adjoint representations of $SU(2)$ come into play.
The Dirac equation for the singlet representation is identical
to \dilatino~and, as stated above, has no normalizable solutions.

In order to find the number of zero--modes of the gaugino in the fundamental
and in the adjoint representation of $SU(2)$ we recall the general formula
for the index, $ind(\fey_V,M,\de M)$, of the
Dirac operator, $\fey_V$, coupled to the spin bundle over the manifold $M$
tensored with a vector bundle $V$. For our purposes, the Yang-Mills connection
on $V$ will be eventually identified with the spin connection.
The formula for $ind(\fey_V,M,\de M)$ reads
$$
\eqalign{
ind(\fey_V,M,\de M)&= {dim V \over 192 \pi^2} \lbrack\int_M tr(R\wedge R)
-\int_{\de M}tr(\theta\wedge R)\rbrack-{1 \over 8 \pi^2}\lbrack\int_M
tr_V(F\wedge F) \cr
& -\int_{\de M} tr_V(\alpha\wedge F)\rbrack-\mezzo
\lbrack\eta(\fey_V,\de M)+h(\fey_V,\de M)\rbrack \cr}
\eqn\indice
$$
In \indice~there are three kinds of contributions. Contributions from
the ``bulk'' which involve only the curvatures $R$ and $F$. Local boundary
contributions, which besides $R$ and $F$ involve the corresponding
second fundamental forms, $\theta$ and $\alpha$, on $\de M$.
Finally there are non-local boundary contributions given in terms of
two invariants, $h(\fey_V,\de M)$ and $\eta(\fey_V,\de M)$.

The $\eta(\fey_V,\de M)$-invariant is the analytic continuation to $s=0$ of the
meromorphic function $\eta(\fey_V,\de M,s)$ defined for $\Re(s)>2$ by
ÿ$$
\eta(\fey_V,\de M,s)=\sum_{\lambda\not= 0} |\lambda |^{-s} sign \lambda
=\sum_{\lambda >0}\lambda^{-s}-\sum_{\lambda <0}(-\lambda)^{-s}
\eqn\etainv
$$
with $\lambda$'s the non--zero eigenvalue of $\fey_V |_{\de M}$.
We remark that $\eta(\fey_V,\de M)$ is left invariant by multiplication
of the $\lambda$'s by arbitrary positive real numbers and hence, in particular,
by a conformal rescaling of the metric on $\de M$.

The ``harmonic correction'' $h(\fey_V,\de M)$ represents the number of
zero eigenvalues of the Dirac operator on the boundary, $\fey_V |_{\de M}$
and coincides with the dimension of the space of the harmonic functions of
the operator $D_V^2$ restricted to the boundary.

Let's start the actual computation with the two ``bulk'' contributions.
The first term in \indice~is immediately evaluated using \hirz.
The calculation of $\int_M tr_V(F\wedge F)$ is reduced to the previous one
by observing that the embedding \embuno~implies
$$
2~tr_{\underline2}F\wedge F = tr R\wedge R
\eqn\identrace
$$
and that for the $SU(2)$ generators in the representation $\underline R$
one has the identity
$$
tr_{\underline R} (T^a T^b) = {2 \over 3}
t(t+1)(2t+1)tr_{\underline2}(T^a T^b)
\eqn\tracce
$$
with $dim \underline R = 2t+1$. Combining these results, one finds
$$
ind(\fey_V,M,\de M)_{bulk}= 3 \lbrack{2t+1 \over 24}-{1 \over 3}
t(t+1)(2t+1)\rbrack
\eqn\bulkindex
$$
We will use this formula for $t=\mezzo,1$ recalling (see \decomposition)
that the adjoint representation of $E(8)$ contains one triplet and $56$
doublets when decomposed with respect to the istantonic $SU(2)$.

As for the boundary contributions, the local terms vanish because by explicit
computation the integrands go to zero, as $r_0$ go to $\infty$, faster
than the integration measure.
We now turn to the non-local boundary terms. As we shall see, the Dirac
operator on the boundary has no zero eigenvalues implying
$$
h(\fey_V,\de M)=0
\eqn\hinv
$$
\REF\pope{C.N.~Pope, J.Phys. {\bf A14} (1981) L133.}
In order to compute the $\eta$-invariant, following [\pope], it is
convenient to separate the Dirac operator on $\de M$
into two terms, a gauge singlet part, in which only the singlet boundary Dirac
operator, $\fey_S |_{\de M}$, appears and a gauge non--singlet part.
The singlet Dirac operator on the boundary has the form
$$
\eqalign{
-i \fey_S |_{\de M} =
-i \t^p e^i_p(\de_i+{1 \over 4}\omega_{i(\de M)}{}^{pq}\t_p\t_q)
\cr}
\eqn\diracbounduno
$$
where $i,p,q=1,2,3$, $a=0,1,2,3$ and the subscript
$(\de M)$ is to remind that the spin connection, $\omega$, is computed on
$\de M$, \ie~on a slice of constant $r=r_0$ (to be sent to $\infty$ at the
end of the calculations).
The appearance of the Pauli matrices, $\t_p$, follows from the last equation
in the formula (A.2).
A semplification arises from the factorization
property of the EH metric on the boundary, which imply the equations
$$
\omega_{(\de M)}{}^{pq}  \equiv \omega_{i(\de M)}{}^{pq} dx^i =
\omega_{\m(M)}{}^{pq} dx^\m |_{\de M}
\eqn\spinbordo
$$
In fact together with the obvious vanishing of the components
$\omega_{i(\de M)}{}^{0p}$, the above formulae allow us to write
$$
\eqalign{
-i \fey_S |_{\de M} =
-i \t^p e^i_p(\de_i+{1 \over 8}\omega_{i(M)}{}^{ab}|_{\de M}\sigma_a
\sigma_b)
\cr}
\eqn\diracbounddue
$$
so that the expression of the Dirac operator on the boundary, $\de M$,
appears as that of a (three--dimensional) Dirac operator in which
the spin connection part is effectively that of the manifold $M$, only
restrited to $\de M$, \ie~computed at $r=r_0$.
To get the explicit expression of \diracbounddue~we first rewrite it
in the form
$$
-i \fey_S |_{\de M} = -i \t^p e^i_p(\de_i+\mezzo({i \over 2}A^k_i\sigma_k))
\eqn\diracop
$$
having used \connmatrix. Then from the identity
$$
{1 \over 4}\t^p e^i{}_p A^k{}_i \sigma_k={1 \over 4} \sum_{i=1}^3 a^{(i)},
\eqn\connectiden
$$
which follows from \connmatrix, and the explicit expressions of the
coefficients, $a^{(i)}$, given in \explicit, \diracop~finally becomes
$$
-i \fey_S |_{\de M} = 2 \pmatrix{L_3 & L_{-}\cr  L_+ & L_3 \cr}+
{3 \over 2}
\pmatrix{ 1 & 0\cr 0 & 1\cr}
\eqn\singletop
$$
where the differential operators $L_\pm$ and $L_3$ are defined in Appendix A.

Using the previous notations, one finds, with a little algebra,
for the boundary Dirac operator in the doublet representation
$$
-i \fey_{\underline 2}|_{\de M}=\pmatrix{-i \fey_S & 0 \cr 0 & -i \fey_S \cr}+
\pmatrix{ \t_3 & \t_1-i\t_2 \cr \t_1+i\t_2 & -\t_3 \cr}
\eqn\doppietto
$$
To lighten the notation
we have dropped the subscript ${\de M}$ on $\fey_S$ inside the matrix
in the r.h.s. of \doppietto.
The very form of $-i \fey_{\underline 2}|_{\de M}$ suggests that we solve the
eigenvalue equation
$$
-i \fey_{\underline 2} |_{\de M} \Psi = \lambda \Psi
\eqn\autovalori
$$
by expanding $\Psi$ in terms of the $SU(2)$ rotation matrices,
$D^l_{n,m}(\theta,\phi,\psi)$, whose definition is recalled in Appendix A.
In this way the infinite dimensional eigenvalue problem for the differential
operator $\fey_{\underline 2}|_{\de M}$ becomes a finite dimensional
eigenvalue problem for a 4x4 matrix with non-vanishing determinant.
This fact implies $\lambda \neq 0$, thus the validity of equation \hinv.

For generic, fixed, values of $l$, $n$ and $m$ (notice that the index $m$
is not acted upon by the operator \doppietto) there are four distinct
eigenvectors that may be written in the form
$$
\Psi^{(i)l}_{n,m}= \pmatrix{c_1^{(i)} D^l_{n-1,m}\cr c_2^{(i)} D^l_{n,m}\cr
c_3^{(i)} D^l_{n,m} \cr c_4^{(i)} D^l_{n+1,m}\cr} \quad\quad i=1,2,3,4
\eqn\eigenvector
$$
with suitable coefficients, $c_k^{(i)}$, whose explicit expression we
will not need here. Given the restrictions on the allowed values of $n$ and $m$
following from the definition of the $D^l_{n,m}$, the form \eigenvector~of
the eigenvectors will retain its validity only if $l\neq 0$, $|m|\leq l$ and
$|n|\leq l-1$, because otherwise some of the entries in \eigenvector~will
loose their meaning. We will discuss separately these limiting cases below.

The four eigenvalues corresponding to the four eigenvectors \eigenvector~are
$$
\eqalign{
\lambda^{(1)l}_{n,m} & = 2l+{5 \over 2}\quad\quad
\lambda^{(3)l}_{n,m} = -2l+\mezzo\cr
\lambda^{(2)l}_{n,m} & = 2l+\mezzo\quad\quad
\lambda^{(4)l}_{n,m} = -2l-{3 \over 2}\cr}
\eqn\eigenvalue
$$
All of them have identical multiplicity, $d$, equal to the product of the
degeneracies of the values of the two ``magnetic'' indices, allowed by the
form \eigenvector~of the eigenvector $\Psi^{(i)l}_{n,m}$,
\ie~$d=d_m d_n= (2l+1)(2l-1)$.

As we mentioned above, besides the eigenvectors and eigenvalues given by
\eigenvector~and \eigenvalue~there are few exceptional cases in which
either $l=0$ or $n$ violates the restriction $|n|\leq l-1$.
For $l\neq 0$ the proper form of the eigenvector is obtained by setting
to zero in the expression \eigenvector~the entries which loose meaning
when $n$ takes the limiting values $n=\pm l$ or $n= \pm (l+1)$.
For $l=0$ all the entries in \eigenvector~should be taken as $\theta,\phi,\psi$
-independent constants. In detail one has

\item{(i)}
$n=l+1$ \nextline
The last three entries in \eigenvalue~have to be set to zero:
1 eigenvector, $\Psi^{l}_{l+1,m}$, with eigenvalue
$2l+{5\over2}$ and multiplicity $d=2l+1$;

\item{(ii)}
$n=-l-1$  \nextline
The first three entries in \eigenvalue~have to be set to zero:
1 eigenvector, $\Psi^{l}_{-l-1,m}$, with eigenvalue
$2l+{5\over2}$ and multiplicity $d=2l+1$;

\item{(iii)}
$n=l$  \nextline
The last  entry in \eigenvalue~has to be set to zero:
3 eigenvectors, $\Psi^{(k)l}_{l,m}, k=1,2,3$, with eigenvalues
$2l+{5\over2}, 2l+{1\over2}, -2l-{3\over2}$, all with multiplicity $d=2l+1$;

\item{(iv)}
$n=-l$  \nextline
The first entry in \eigenvalue~has to be set to zero:
3 eigenvectors, $\Psi^{(k)l}_{-l,m}, k=1,2,3$, with eigenvalues
$2l+{5\over2}, 2l+{1\over2}, -2l-{3\over2}$, all with multiplicity $d=2l+1$;

\item{(v)}
$l=0$  \nextline
$2$ constant eigenvectors with eigenvalues ${5 \over 2}, -{3 \over 2}$
and multiplicity $d = 3, 1$ respectively.

We are now in position to compute $\eta(\fey_{\underline 2},\de M)$ using
formula \etainv.
We first notice that the identification of antipodal points in the
EH manifold forces the sum over the angular momentum, $l$,
implicit in \etainv, to run only over integral values.
The contribution from $l=0$ is easily evaluated and gives
$$
\eta(\fey_{\underline 2},\de M,s)_{l=0} =
 {3 \over ({5 \over2})^s}-{1 \over ({3 \over2})^s}
\eqn\etazero
$$
A little more work is necessary to compute the contribution from $l\neq 0$
which turns out to be
$$
\eqalign{
\eta(\fey_{\underline 2},\de M,s)_{l\neq 0}&=
{ 1 \over 2^s}\sum_{l=1}^\infty \lbrack
{(2l+1)(2l+3) \over (l+{5 \over 4})^s}+{(2l+1)^2 \over (l+{1 \over 4})^s}\cr
&-{(2l+1)^2 \over (l+{3 \over 4})^s}
-{(2l+1)(2l-1) \over (l-{1 \over 4})^s} \rbrack \cr
&={ 1 \over 2^s}\lbrack 4\z (s-2,{9 \over 4})-2\z (s-1,{9 \over 4})
-{3 \over 4} \z (s,{9 \over 4}) +{3 \over 4}\z (s,{3 \over 4})\cr
&+4 \z (s-2,{5 \over 4})+2\z (s-1,{5 \over 4})
+{1 \over 4}\z (s,{5 \over 4})-4\z (s-2,{7 \over 4})\cr
&+2\z (s-1,{7 \over 4})
-{1  \over 4}\z (s,{7 \over 4})-4 \z (s-2,{3 \over 4})-2\z (s-1,{3 \over 4})
\rbrack \cr}
\eqn\etaesse
$$
The sum of the various Riemann zeta functions in \etaesse~is evaluated
by using the known relation between the Riemann zeta functions and the
Bernoulli polynomials, which reads  $\zeta(-n,q)=-{B_{n+1}(q) \over n+1}$.
Adding \etazero~to \etaesse~and taking the limit $s\rightarrow 0$,
one gets for the $\eta$-invariant
$$
\eta(\fey_{\underline 2},\de M)=\eta(\fey_{\underline 2},\de M)_{l=0}+
\eta(\fey_{\underline 2},\de M)_{l\neq 0}= 2-{5 \over 2}=-\mezzo
\eqn\etatotale
$$
Inserting \bulkindex, \hinv~and \etatotale~in \indice, we finally find
$$
ind(\fey_{\underline2},M,\de M)=-1
\eqn\totalindex
$$
The equality of $ind(\fey_{\underline2},M,\de M)$ with the value \tauEH~of
the Hirzebruch signature
can be traced back to the observation that, for the doublet representation,
isospin $\mezzo$ and spin $\mezzo$  can be combined together to get a total
spin of 1 or 0. We can then compute
the index of the Dirac operator in the doublet representation by relating
it to a formula (to be found in [\egh]) that gives the index of the generalized
Dirac operator acting on ``spinorial forms'' of left--handed spin
${m \over 2}$ and right--handed spin ${n \over 2}$.
These ``spinorial forms'' are sections of
a spin bundle which is the symmetric tensor product of $m$ copies of the
elementary spin bundle of a left--handed spinor and of $n$ copies of the
elementary spin bundle of a right--handed spinor.
In the EH case this formula reduces to
$$
\eqalign{
ind (\fey_{({m \over 2},{n \over 2})})&={1 \over 32}(m+1)(n+1)\lbrack (-1)^m
-(-1)^n\rbrack \cr
&+{(m+1)(n+1) \over 240} \lbrack m(m+2)(3m^2+6m-14) \cr
&-n(n+2)(3 n^2+6n-14) \rbrack \tau \cr}
\eqn\indexancora
$$
A straightforward application of \indexancora~leads indeed to the results
$ind (\fey_{(1,0)})=\tau$ and $ind (\fey_{(0,0)})=0$.

The above observation will be exploited in the next section, where we will
find the explicit form of the (left--handed) doublet zero--mode.
Since no right--handed zero--modes are expected in a self--dual background
such as the heterotic EH instanton, the unique doublet zero--mode should be
identified with the one obtainable from the (unique,
$|\t_{\scriptscriptstyle{EH}}|=1$)
normalizable closed self--dual two--form existing on the EH manifold.

The calculation of $ind(\fey_{\underline3},M,\de M)$ can be performed
in complete analogy with the previous case. In order to
separate the angular momentum dependence, instead of using the standard
form of the generators in the adjoint representation,
$(T^i)_{jk}=-i (\varepsilon^i)_{jk}$, it turns out to be more convenient
to diagonalize $T^3$ and write
$$
T^3=\pmatrix{1 & 0 & 0\cr 0 & 0 & 0 \cr 0 & 0 & -1 \cr},
T^1={1 \over \sqrt{2}}\pmatrix{0 & -1 & 0 \cr -1 & 0 & 1 \cr 0 & 1 & 0\cr},
T^2={1 \over \sqrt{2}}\pmatrix{0 & i & 0 \cr -i & 0 & -i \cr 0 & i & 0\cr}
\eqn\repapp
$$
The Dirac operator which will now act on a spinor of the form
$\vec \Psi = (\Psi^1,\Psi^2,\Psi^3)$ with $\Psi^k, k=1,2,3$,
two-component spinors, takes the form
$$
-i \fey_{\underline3} |_{\de M} = \pmatrix{ -i \fey_S & 0 & \cr 0 & -i
\fey_S & 0 \cr 0 & 0 & -i \fey_S \cr}+2 \pmatrix{2\t_3 & -\t_- & 0 \cr
-\t_+ & 0 &  \t_- \cr
0 & \t_+ & -2 \t_3 \cr}
\eqn\diratrip
$$
where $\t_+=\sqrt{2}(\t_1+i\t_2), \t_-=\sqrt{2}(\t_1-i\t_2)$.

For generic, fixed, values of $l$, $m$ and $n$, with $l\neq 0$,
$|m|\leq l$ and $-l+1\leq n \leq l-2$ (seven special
cases will have to be considered separately), there are six distinct
eigenvectors. They have the form
$$
\Psi^{(i)l}_{n,m}= \pmatrix{c_1^{(i)} D^l_{n-1,m}\cr c_2^{(i)} D^l_{n,m}\cr
c_3^{(i)} D^l_{n,m} \cr c_4^{(i)} D^l_{n+1,m}\cr c_5^{(i)} D^l_{n+1,m}\cr
c_6^{(i)} D^l_{n+2,m}\cr} \quad\quad  i=1,\dots ,6
\eqn\eigenvectortri
$$
with suitable coefficients, $c_k^{(i)}$, whose explicit expression we
will not need here. The corresponding six eigenvalues are
$$
\eqalign{
\lambda^{(1)l}_{n,m} & = 2l+{7 \over 2}\quad\quad
\lambda^{(4)l}_{n,m} = -2l-{5 \over 2}\cr
\lambda^{(2)l}_{n,m} & = 2l+{3 \over 2}\quad\quad
\lambda^{(5)l}_{n,m} = -2l-{1 \over 2}\cr
\lambda^{(3)l}_{n,m} & = 2l-{1 \over 2}\quad\quad
\lambda^{(6)l}_{n,m} = -2l+{3 \over 2}\cr}
\eqn\eigenvaluetri
$$
all with identical multiplicity, $d$, equal to the product of the
degeneracies of the values of the two ``magnetic'' indices, allowed by
the form \eigenvectortri~of the eigenvector $\Psi^{(i)l}_{n,m}$,
\ie~$d=d_m d_n= (2l+1)(2l-2)$.

The seven special cases in which some of the entries of \eigenvectortri~loose
meaning are separately listed below.

\item{(i)}
$n=l+1$  \nextline
The last five entries in \eigenvectortri~have to be set to zero:
1 eigenvector, $\Psi^{l}_{l+1,m}$, with eigenvalue
$2l+{7 \over2}$ and multiplicity $d=2l+1$;

\item{(ii)}
$n=-l-2$  \nextline
The first five entries in \eigenvectortri~have to be set to zero:
1 eigenvector, $\Psi^{l}_{-l-2,m}$, with eigenvalue
$2l+{7 \over2}$ and multiplicity $d=2l+1$;

\item{(iii)}
$n=l$  \nextline
The last three enties in \eigenvectortri~have to be set to zero:
3 eigenvectors, $\Psi^{(k)l}_{l,m}, k=1,2,3$, with eigenvalues
$2l+{7\over2}, 2l+{3\over2}, -2l-{5\over2}$, all with multiplicity $d=2l+1$;

\item{(iv)}
$n=-l-1$  \nextline
The first three entries in \eigenvectortri~have to be set to zero:
3 eigenvectors, $\Psi^{(k)l}_{-l-1,m}, k=1,2,3$, with eigenvalues
$2l+{7\over2}, 2l+{3\over2}, -2l-{5\over2}$, all with multiplicity $d=2l+1$;

\item{(v)}
$n=l-1$  \nextline
The last entry in \eigenvectortri~has to be set to zero:
5 eigenvectors, $\Psi^{(k)l}_{l-1,m}, k=1,\dots ,5$, with eigenvalues
$2l+{7\over2}, 2l+{3\over2}, 2l-{1\over2}, -2l-{1\over2}, -2l-{5\over2}$,
all with multiplicity $d=2l+1$;

\item{(vi)}
$n=-l$ \nextline
The first entry in \eigenvectortri~has to be set to zero:
5 eigenvectors, $\Psi^{(k)l}_{-l,m}, k=1,\dots ,5$ with eigenvalues
$2l+{7\over2}, 2l+{3\over2}, 2l-{1\over2}, -2l-{1\over2}, -2l-{5\over2}$,
all with multiplicity $d=2l+1$;

\item{(vii)}
$l=0$  \nextline
$2$ constant eigenvectors with eigenvalues ${7 \over 2}, -{5 \over 2}$
and multiplicity $d = 4, 2$ respectively.

Putting all these results together, we get $\eta(D_{\underline3},\de M)=
{3 \over 4}$ and finally
$$
ind(\fey_{\underline3},M,\de M)=ind(\fey_{\underline3},M,\de M)_{bulk}-
{3\over 8}= -{45 \over 8} -{3 \over 8}=-6
\eqn\indextotal
$$
Exactly six left--handed zero--modes should be found for the triplet Dirac
operator in the heterotic EH instanton, because no right--handed zero--modes
are expected in a self--dual background. Their explicit expressions as well
as those of all the bosonic and fermionic zero--modes will be given in the
next section.

\chapter{Zero--modes and Their Geometrical Interpretation}
In this section we will present the explicit form of the zero--modes of the
wave operators that control the quadratic fluctuations of the lagrangian
\boselagrangian, \fermilagrangian, expanded around the heterotic EH
instantonic classical solution we have discussed in sect.~2.3.

These calculations are propaedeutical to any computational scheme which is
intended to extend the instanton calculus, developed in globally
supersymmetric theories, to the case of supergravity [\taylor].
With an eye to future applications [\noi] to semi-classical instanton
calculations, we will give the explicit expression of the fermionic and bosonic
zero--modes in the the  gauge-fixed form most appropriate for this purpose.
The natural choice for the gauge-fixing of the functional
integral [\tof, \yaf, \konuno, \noi] induce the following gauge conditions
on the gauge vector, graviton and gravitino zero-modes:
\item{(i)}
background transversality for the gauge zero--modes;
\item{(ii)}
background transversality and tracelessness for the graviton zero--modes;
\item{(iii)}
background $\gamma$-tracelessness for the gravitino zero--modes.

As is well known,
zero--modes of the bosonic fields are in one to one correspondence with the
collective coordinates or moduli of the classical solution. They correspond
to explicit parameters appearing in the instantonic solution or to
classical symmetries which are broken by it. Zero--modes of the fermionic
fields are associated to the superpartners of the bosonic collective
coordinates. We present in the Table a summary of the results on zero--modes
and
collective coordinates which we are now going to derive in detail.

$$\vbox{%
\newbox\mystrutbox
   \setbox\mystrutbox=\hbox{\vrule height15pt depth7pt width0pt}
   \def\mystrut{\relax\ifmmode\copy\mystrutbox\else\unhcopy\mystrutbox\fi}
\def\doublebox#1#2#3{$\vcenter{\vskip 5 pt
   \hbox to #1 truecm {\hss\strut#2\hss}
   \hbox to #1 truecm {\hss\strut#3\hss}
   \vskip 5pt}$}
\def\triplebox#1#2#3#4{$\vcenter{\vskip 5 pt
   \hbox to #1 truecm {\hss\strut#2\hss}
   \hbox to #1 truecm {\hss\strut#3\hss}
   \hbox to #1 truecm {\hss\strut#4\hss}
   \vskip 5pt}$}
\rm \tabskip=0pt \offinterlineskip
\halign {\mystrut#&%
\vrule#&\quad\hfil#\hfil\quad&%
\vrule#&\quad\hfil#\hfil\quad&%
\vrule#&\quad\hfil#\hfil\quad&%
\vrule#&\quad\hfil#\hfil\quad&%
\vrule#&\quad\hfil#\hfil\quad&%
\vrule#\cr
\noalign{\hrule}
& & \multispan3 \hfil Field \hfil
& & $n_{\scriptscriptstyle{0}}$
& & \doublebox{2}{Collective}{coordinates}
& & Broken symmetry & \cr
\noalign{\hrule}
& & $ \rm B_{\mu \nu}$ & & axion  & & 1 & & $b$
& & Peccei-Quinn shift & \cr
\noalign{\hrule}
& & $ \psi^{\alpha}_{\mu} $ & & gravitino & & 2 & & $ \eta^{\alpha} $
& & Supersymmetry & \cr
\noalign{\hrule}
& &  & &  & &  & & $a$ & & Dilatations & \cr
\omit & & $ \rm g_{\mu \nu} $ & & metric & & 3 & & \omit \hrulefill & & \omit
\hrulefill & \cr
& &  & &  & &  & & $ \beta_1 , \beta_2 $ & &
\doublebox{2}{$ \rm SU(2)_L / U(1)_L $}{rotations} & \cr
\noalign{\hrule}
& & $ A^i_{\mu} $
& & \doublebox{2}{gauge vector}{(triplet)}
& & 12 & & $ \gamma_1 , \ldots , \gamma_{12} $
& & \triplebox{3.2}{Global SU(2) gauge}
                 {rotations $=$ tangent bundle} {deformations} & \cr
\noalign{\hrule}
& & $ \chi^i_{\alpha}$
& & \doublebox{2}{gaugino}{(triplet)}
& & 6 & & $ \alpha_1 , \ldots , \alpha_6 $
& & $ \gamma $ superpartners  & \cr
\noalign{\hrule}
& & $ A^r_{\mu} $
& & \doublebox{2}{gauge vectors}{(doublets)}
& & 112 & & $ \theta_1 , \ldots, \theta_{112}$
& & \doublebox{4}{Global $ \rm E_8 / SU(2) \otimes E_7 $}
                 {gauge rotations} & \cr
\noalign{\hrule}
& & $ \chi^r_{\alpha}$
& & \doublebox{2}{gaugino}{(doublets)}
& & 56 & & $ \delta_1 , \ldots, \delta_{56} $
& & $ \theta $  superpartners & \cr
\noalign{\hrule}
\noalign{\bigskip}
& \multispan{11} \hfil Zero--modes and collective coordinates for heterotic
Eguchi--Hanson Instanton \hfil \cr }}$$

Following [\hp, \konuno], the construction of the gravitino and graviton
zero--modes can be easily carried out, by exploiting
properties of the EH manifold that are listed below.

\item {(i)}
In the EH manifold there exists a closed self--dual two--form whose explicit
components, in the frame specified by the tetrad \tetrad, are
$$
B_{ab} = \eta^3_{ab} \ar
\eqn\twoform
$$

\item {(ii)}
The self--duality of the spin connection and the vanishing of the
curvature scalar of the EH metric imply that both the square of the
Rarita-Schwinger operator and the Liechnerowicz operator (the latter governs
the propagation of the fluctuations of the metric) are identical
in the chosen background gauge to the
Laplace-Beltrami operator acting on self--dual two--forms [\duffdue].

\item {(iii)}
In any self--dual background there always exist two covariantly constant
right--handed spinors, $\bar\epsilon^{(r)}_{\adot}, r=1,2$. Actually in the
EH background they are just constant spinors.

Using properties (i) to (iii), we conclude that the spinor components
of the two spin ${3 \over 2}$ zero--modes should have the expression

$$
\psi^{(r)}_{\a\b{\adot}} = B_{ab} \s^{ab}{}_{\a\b}
\bar\epsilon^{(r)}_{\adot},\quad r=1,2
\eqn\psizeromodes
$$
These zero--modes can be rewritten as Rarita-Schwinger spinors
in the form
$$
\psi^{(r)}_\mu = D_\mu \eta^{(r)} - {1\over 4} \s_\m \> \fey\eta^{(r)}
\eqn\etazeromodes
$$
having introduced the left--handed spinor $\eta^{(r)}= u \eta^{(r)}_0$ with
$\eta^{(r)}_0= i\s_0 \bar\epsilon^{(r)}$.
The second term in the r.h.s. of \etazeromodes~has been added to insure
the validity of the gauge condition $\gamma_\m \psi^\m = 0$. It can be
interpreted as a global super Weyl rescaling, since
$({\fey\eta^{(r)}})_{\adot} = \bar\epsilon^{(r)}_{\adot} = const$.

The above construction is telling us that the existence of two gravitino
zero--modes should be ascribed to the lack of invariance of the EH background
under the two global supersymmetry transformations induced by the two
left--handed spinors, $\eta^{(r)}$.

The three independent spin 2 zero--modes can be similarly constructed by
performing a further supersymmetry transformation on \etazeromodes. One obtains
$$
h^{(I)}_{\a\b\adot\bdot} =
\psi^{(r)}_{\a\b\adot} \bar\epsilon^{(s)}_{\bdot} C^{(I)}_{rs},
\quad  I=0,1,2
\eqn\hzeromodes
$$
where we have defined the Clebsch-Gordan coefficients
$$
C^{(0)}_{rs}  = \d_{rs}\quad
C^{(1)}_{rs}  = (\t_3)_{rs}\quad
C^{(2)}_{rs}  = (\t_1)_{rs}
\eqn\clebsch
$$
In the more standard tensor notation we can write
$$
h^{(I)}_{\mn} = \psi^{(r)}_\m \s_\n \bar\epsilon^{(s)} C^{(I)}_{rs}
+ \psi^{(r)}_\n \s_\m \bar\epsilon^{(s)}  C^{(I)}_{rs}
\eqn\hsusy
$$
Substituting \etazeromodes~in \hsusy~and
using again the fact that $\bar\epsilon$ is covariantly constant,
the expression \hsusy~can be cast in the form of an infinitesimal
diffeomorphism accompanied by a global Weyl rescaling. One can in fact write
$$
\eqalign{
h^{(I)}_{\m\n}& =  C^{(I)}_{rs} \eta^{(r)}_0\s_a\bar\epsilon^{(s)}
(\nabla_\m (u e^a_\n)+\nabla_\n (u e^a_\m)-
\mezzo \nabla\cdot (u e^a)g_{\m\n})\cr
&= \nabla_\m \zeta^{(I)}_\n + \nabla_\n \zeta^{(I)}_\m
- \mezzo (\nabla\cdot\zeta^{(I)}) g_{\mn} \cr}
\eqn\hdiff
$$
with
$$
\zeta^{(I)}_\m = \eta^{(r)}\s_\m\bar\epsilon^{(s)} C^{(I)}_{rs}
\eqn\zetadef
$$
In \hdiff~the derivative $\nabla_\m$ contains only the
Christoffel connection, while $D_\m$ is fully covariant. Actually the last
term in \hdiff~in non-zero only for $I=0$ because, as it can be easily shown,
$\nabla\cdot\zeta^{(1,2)}=0$, while $\nabla\cdot\zeta^{(0)}=constant$.

With a suitable interpretation of the infinitesimal diffeomorphisms
$\zeta^{(I)}_\m$ these zero--modes
may be related to the lack of invariance of the EH metric
under dilatations and under the two rotations in the coset $SU(2)_L/U(1)_L$
(the $U(1)_L$ quotient factor is present
because the corresponding generator, $\xi_{(3)}$, given in \killinguno, is
a Killing vector of the EH metric).

This can be explicitely seen by separating
the diffeomorphisms in \hdiff~into a ``rigid'' part,
$\xi_{(I)}^{\m}$, related to dilations ($I=0$) or to the two
rotations in the coset $SU(2)_L/U(1)_L$ ($I=1,2$), and a ``gauge''
transformation part, $\Lambda_{(I)}^{\m}$, whose presence is necessary
to ensure the transversality of the corresponding zero--modes, $h^{(I)}_{\mn}$.

The infinitesimal diffeomorphism related to dilatations can be split as follows
$$
\z_{(0)}^{\m} = ({u\over r})^2 x^{\m} = x^{\m} - \ar x^{\m}
\equiv \xi_{(0)}^{\m}+\Lambda_{(0)}^{\m}
\eqn\hdil
$$
where the first term, $\xi_{(0)}^{\m}=x^{\m}$, clearly corresponds to a global
rescaling. Similarly the other two zero--modes are associated to the global
rotations induced by the two vector fields
$$
\zeta_{(i)}^{\m} = u e_{(i)}^{\m}
= \eta_i{}^\m{}_\n x^\n + ({u\over r} -1)\eta_i{}^\m{}_\n x^\n
\equiv \xi_{(i)}^{\m}+\Lambda_{(i)}^{\m}
\eqn\hrot
$$
with $i=1,2$ and $\eta_i{}^\m{}_\n$ the 't Hooft symbols.

As is well known [\prasad], the metric \ehmetric~is obtained from (B.1)
with a change of variables which puts
the (dipole with) endpoints $\vec x_1, \vec x_2$ along a preferred axis.
The length of this dipole is related to the modulus $a$.
The effect of the rotations $\xi_{(1,2)}^\m$ is to change the alignment of the
dipole and that of the dilatation $\xi_{(0)}^{\m}$ to rescale the distance
$|\vec x_1 - \vec x_2|$.

A more suggestive form of the graviton zero--modes, $h^{(I)}_{\mn}$, can
be obtained by noticing that the ``rigid'' transformations considered above
can be interpreted as Lie derivatives along the congruences of the vector
fields $\z_{(I)}^\m \de_\m$. The curves of the congruences are generated by the
exponential maps of these vector fields and can be parametrized by the moduli
of the Eguchi--Hanson instanton,
$a$ for the dilatations and $\beta_1, \beta_2$ for the two rotations.

The Lie derivative represents a diffeomorphism along the congruence,
\ie~an infinitesimal change of the moduli, and can thus be also seen as
the derivative of the metric with respect to the moduli themselves.

{}From the explicit form of the EH metric we have the formulae
$$\eqalign{
{\de g_{\m\n}\over \de \beta_i} & = \xi_{(i)}^\lambda \de_{\lambda}g_{\m\n}+
g_{\m\lambda}\de_{\n}\xi^{\lambda}_{(i)}+
g_{\n\lambda}\de_{\m}\xi^{\lambda}_{(i)}\cr
& = \nabla_{\m}\xi_{(i)\n} + \nabla_{\n}\xi_{(i)\m},\quad i=1,2\cr}
\eqn\liederivativei
$$

$$\eqalign{
{\de g_{\m\n}\over \de a}| & = \xi_{(0)}^\lambda \de_{\lambda}g_{\m\n}+
g_{\m\lambda}\de_{\n}\xi^{\lambda}_{(0)}+
g_{\n\lambda}\de_{\m}\xi^{\lambda}_{(0)} - \mezzo (\nabla\cdot\xi_{(0)})
g_{\mn} \cr
& = \nabla_{\m}\xi_{(0)\n} + \nabla_{\n}\xi_{(0)\m} - \mezzo
(\nabla\cdot\xi_{(0)}) g_{\mn}\cr}
\eqn\liederivativezero
$$
The $|$ in the l.h.s. of \liederivativezero~is to remember that the derivative
with respect to $a$ must be taken by keeping constant the determinant of the
metric. This constraint is responsible for the appearance of the last extra
term in the r.h.s. of \liederivativezero. The condition that the determinant of
the metric should remain a constant comes, in turn, from the requirement that
the fluctuation of the metric should be traceless.

Using equations \liederivativei~and \liederivativezero, one can finally write
$$\eqalign{
h^{(0)}_{\mn} &={\de g_{\mn} \over \de a}| +
\nabla_\m \Lambda^{(0)}_{\n} + \nabla_\n \Lambda^{(0)}_{\m}\cr
h^{(1)}_{\mn} &={\de g_{\mn} \over \de \beta_1}+
\nabla_\m \Lambda^{(1)}_{\n} + \nabla_\n \Lambda^{(1)}_{\m}\cr
h^{(2)}_{\mn} &={\de g_{\mn} \over \de \beta_2}+
\nabla_\m \Lambda^{(2)}_{\n} + \nabla_\n \Lambda^{(2)}_{\m}\cr}
\eqn\hzeromodi
$$
In this way each graviton zero--mode is expressed as a
sum of two terms. The first is the derivative of the EH metric with respect to
a collective coordinate, the second is a gauge transformation
necessary to make the zero--mode to obey the same gauge condition as the
metric, \ie~to make it transverse (tracelessness is automatic from \hdiff).

\REF\popeuno{C.N.~Pope and A.L.~Yuille, Phys.Lett. {\bf 78B} (1978) 424;
\nextline C.N.~Pope, Nucl.Phys. {\bf B141} (1978) 432.}
We now turn to the study of the gaugino zero--modes, starting with the doublet
case (for other gravitational instantons, analogous computations can be found
in [\popeuno]).
Using \connmatrix, we can write the Dirac equation in the doublet
representation in the form
$$
-i\fey_{\underline2}\chi = -i \s^a{}_{\a{\adot}}\de_a\chi^{\a r}+
{i \over 2} a^{(k)}(\delta_{\adotb}
\delta^r{}_s+(\t^k)^r{}_s(\t^k)_{\adotb})\chi^{\b s}=0
\eqn\zerodopp
$$
with $a^{(k)}$ defined in \explicit.
It is easy to see that an ansatz of the form
$$
\chi^{\a r} = \pmatrix {P_1 D^l_{n-1,m} & Q_1 D^l_{n,m}\cr
P_2 D^l_{n,m} & Q_2 D^l_{n+1,m}\cr}
\eqn\gauginoansa
$$
with $P_1, P_2, Q_1, Q_2$ purely radial functions \foot{To lighten the
notation we have dropped all the angular momentum indices from the functions
$P_1, P_2, Q_1, Q_2$} and $D^l_{n,m}$ the $SU(2)$ rotation matrices,
leads to the following system of first order differential equations

$$
\eqalign{
\lbrack {u \over r} {d \over dr}+{2 (n+1) \over u}\rbrack P_1+{2 \over r}
a_{nl} P_2 &=0 \cr
\lbrack {u \over r} {d \over dr}+{2u \over r^2}-{2 n \over u}\rbrack P_2+
{2 \over r} a_{nl} P_1 +{2 u \over r^2} Q_1 &=0 \cr
\lbrack {u \over r} {d \over dr}+{2u \over r^2}+{2 n \over u}\rbrack Q_1+
{2 \over r} b_{nl} Q_2 +{2 u \over r^2} P_2 &=0 \cr
\lbrack {u \over r} {d \over dr}-{2 (n-1) \over u}\rbrack Q_2+{2 \over r}
b_{nl} Q_1 &=0 \cr}
\eqn\zerodoppuno
$$
where $a_{nl}=\sqrt{(l+n)(l-n+1)}, b_{nl}=\sqrt{(l-n)(l+n+1)}$.

To solve the system \zerodoppuno~we recall that the result of the
index theorem \totalindex~was $ind(\fey_{\underline2},M,\de M)=-1$,
leading us to the conclusion that in the doublet case
there is just one (left--handed) gaugino zero--mode,
since no right--handed zero--modes are expected in a self--dual
background. We are thus led to take $l=0$ in the ansatz \gauginoansa.
With this choice the matrix \gauginoansa~becomes off--diagonal.
Since furthermore $a_{00} = b_{00} = 0$, the system \zerodoppuno~is
immediately solved, giving
$$
\chi^{\a r}=({ a \over r})^4\pmatrix{ 0 & 1\cr 1 & 0\cr}^{\a r}
\eqn\zeromoddue
$$
This formula is very much reminiscent of the form of
the harmonic two--form \twoform. In fact there is another, more
geometrical, way to determine the solution \zeromoddue, which we now
want to explain.
Combining isospin and spin indices, one can separate the symmetric and the
antisymmetric part of $\chi$ by writing
$$
\chi^{\a r}= U\varepsilon^{\a r}+(V_{ab}\s^{ab})^{\a r}
\eqn\chiseparato
$$
where $V_{ab}$ is self--dual. The Dirac equation becomes
$$
\bar\s^\m \lbrack \varepsilon D_\mu U+\s^{ab}
D_\mu V_{ab}\rbrack =0
\eqn\altzero
$$
and thanks to the completeness of the $\bar\s^\m$ matrices, one gets
$$
dU + *dV = 0
\eqn\ciccio
$$
Acting with $d$ or $*d*$ allows to derive from \ciccio~two separate equations.
The first one, $dU=0$, has the only
solution $U=constant$, which is not normalizable.
The second one, $*dV=0$ or, equivalently, $dV=0$ (since $V_{ab}$ is
self--dual) implies that $V$ is a closed self--dual two--form. As there exists
only one (normalizable) self--dual two--form on the EH manifold, $V_{ab}$
must coincide with
$B_{ab}$ of equation \twoform.
This procedure is very general and can be used to find the zero--modes
of the Dirac operator in the doublet representation of the gauge group,
in any self--dual background, whenever spin and gauge connections are
identified.

We now discuss the Dirac equation for the triplet. Due to the lack
of a clear geometrical interpretation, this case will turn out to
be much more involved than the previous one.

The triplet Dirac operator has the form
$$
-i\fey _{\underline 3}=
\pmatrix{-i\fey_S & 0 & 0\cr 0 & -i\fey_S & 0 \cr 0 & 0 & -i\fey_S \cr}
+\pmatrix{B\t_3 & -A\t_- & 0 \cr -A\t_+ & 0 & A\t_- \cr
0 & A\t_+ & -B\t_3 \cr}
\eqn\tuttodiractre
$$
where $A=a^{(1)}=a^{(2)}$, $B=a^{(3)}$. With an eye to the discussion
of the eigenvalue problem for the boundary triplet Dirac operator in sect.~3,
we look for zero--modes of the form
$$
\chi^{\a i}= \pmatrix{R^1 D^l_{n-1,m},
R^2 D^l_{n,m}, S^1 D^l_{n,m}, S_2 D^l_{n+1,m},
T^1 D^l_{n+1,m}, T^2 D^l_{n+2,m}}
\eqn\chitre
$$
with $R^1, R^2, S^1, S^2, T^1, T^2$ purely radial functions
\foot{To lighten the notation we have dropped all angular momentum indices
from the functions $R^1, R^2, S^1, S^2, T^1, T^2$} and
$D^l_{n,m}$ the $SU(2)$ rotation matrices.

With the ansatz \chitre~the Dirac equation, $-i\fey _{\underline 3}\chi=0$,
becomes
$$
\eqalign{
(D+{2 \over u}L_3) R^1+{2 \over r} L_- R^2+B R^1 &=0 \cr
(D-{2 \over u}L_3) R^2+{2 \over r} L_+ R^1-B R^2-\sqrt{2} A S^1 &=0 \cr
(D+{2 \over u}L_3) S^1+{2 \over r} L_- S^2-\sqrt{2} A R^2 &=0 \cr
(D-{2 \over u}L_3) S^2+{2 \over r} L_+ S^1+\sqrt{2} A T^1 &=0 \cr
(D+{2 \over u}L_3) T^1+{2 \over r} L_- T^2-B T^1+\sqrt{2} A S^2 &=0 \cr
(D-{2 \over u}L_3) T^2+{2 \over r} L_+ T^1+ B T^2 &=0 \cr}
\eqn\sistema
$$
where we have introduced the definition
$$
D={u \over r}{d \over dr}+{2 \over u}+ {u \over r^2}
\eqn\deriv
$$

This is a linear system of six Fuchsian differential equations, equivalent
to a Fuchsian differential equations of the sixth order, with three
regular singular points for each one of the radial functions appearing in the
ansatz \chitre.
For special forms of the coefficients, a solution of a differential equation
of the above type can be expressed in terms of the generalized hypergeometric
series. This turns out not to be our case.
We are thus forced to solve \sistema~by brute force, by trying the lowest
possible values of the quantum numbers $l,n,m$, in the hope of successively
reducing the number of independent equations in the system.
We start with the choice $l=n=m=0$. In this case the equations decouple
and one can easily find
a solution which, however, turns out to be non-normalizable.
Next we  examine the case $l=1$ with $-3\leq n \leq 2, -1\leq m \leq 1$.
The values of $n$ we have given are the only ones compatible with the
assignment of angular momentum quantum numbers made in \chitre.
The six resulting possible cases are pairwise symmetric and give rise to the
following three situations.
\item{(i)}
$n=-3, 2$\nextline
The solution has only one non-zero component and it is
not normalizable.
\item{(ii)}
$n=-2, 1$\nextline
The solution has three non-zero components and it is
not normalizable.
\item{(iii)}
$n=-1, 0$\nextline
The solution has five non-zero components and for each value of
$n$ there are three (in correspondence with the three possible values of $m$)
linearly independent normalizable solutions.

In order to explicitely find these solutions it is convenient
to rewrite the system \sistema~as a fifth order differential equation
by solving for one of the five non-zero components of \chitre.
Fortunately three solutions of the resulting Fuchsian equation
may be found with the ansatz: $x^\a (x-1)^\b$, where $x=\ar$ and $\a, \b$
are the exponents governing the behaviour of the solution around
the singularities. Knowing three solutions it is possible to reduce the
original fifth order differential equation to a second order differential
equation of the known hypergeometric type and find the last two
independent solutions. For each $n$ and $m$ one can construct five
independent solutions of \sistema~and it just happens that only one
linear combination of them is normalizable. Explicitely the six normalizable
solutions are ($n = -1, 0; m = -1, 0, 1$)

$$
\chi^{(\downarrow)}_m = \pmatrix{0\cr
-x^{{3 \over 4}}(1+2\sqrt{x})D^1{}_{-1,m}\cr
x^{{5 \over4}}D^1{}_{-1,m}\cr
-{1 \over \sqrt{2}} x^{{3 \over4}}\sqrt{1-x} D^1{}_{0,m}\cr
-{x^{{5 \over 4}}\sqrt{2(1-x)} \over 1+\sqrt{x} } D^1{}_{0,m} \cr
{x^{{3 \over4}}(1-\sqrt{x} ) \over (1+\sqrt{x}) }D^1{}_{1,m} \cr} \quad
\chi^{(\uparrow)}_m = \pmatrix{
-{x^{{3 \over4}}(1-\sqrt{x})  \over ( 1+\sqrt{x}) }D^1{}_{1,m} \cr
{x^{{5 \over 4}}\sqrt{2(1-x)} \over 1+\sqrt{x} } D^1{}_{0,m} \cr
-{1 \over \sqrt{2}} x^{{3 \over4}}\sqrt{1-x} D^1{}_{0,m}\cr
x^{ 5 \over 4}D^1{}_{-1,m} \cr
x^{{3 \over4}}(1+2\sqrt{x})D^1{}_{-1,m} \cr
0 \cr}
\eqn\truemodes
$$
This result is in agreement with the value of the index of the Dirac operator
computed at the end of sect.~3.

We conclude this section with a discussion on the relation between the
number of triplet zero--modes and the number of moduli of an $SU(2)$
gauge instanton over the EH manifold. First of all
notice that the number of zero--mode fluctuations of the gauge field is
twice the (absolute) value of the index \indextotal~for the triplet Dirac
operator.
This is a direct consequence of the already mentioned existence
in self--dual backgrounds, such as the heterotic EH instanton, of two unbroken
right--handed supersymmetries. Thanks to them, two zero--modes of the gauge
field, $\d A_\m^{i(r)}$,
can be generated out of each fermion zero--mode, $\chi^i_\a$, by writing
$$
\d A_\m^{i(r)}= \bar\epsilon^{(r)}_{\adot} \bar\s_\m^{\adot\a}\chi^i_\a
\eqn\azeromodes
$$
Thus the six zero--modes of the gaugino in the adjoint representation
of the istantonic $SU(2)$ generate twelve zero--modes for quadratic
operator of the gauge field fluctuactions around the hetrotic EH background.
At this point the issue is to find a geometric interpretation for
the corresponding 12 ``collective coordinates" of the instantonic gauge
connection.
\REF\wittendt{E.~Witten, Nucl.Phys. {\bf B268} (1986) 79.}
Following [\wittendt] and the discussion in Appendix B \foot{We
thank D.~Anselmi, R.~D'Auria and P.~Fre' for an enlightening discussion
on this point.}, we are led to identify these collective coordinates with the
deformations of the tangent bundle in the adjoint
representation of the structure group of the bundle, which in this case is
$SL(2,\complex)$ (\ie~the complexification of $SU(2)$).
The relevant cohomology group is $H^1(EndT)$, consisting in the holomorphic
(0,1)--forms with values in the adjoint representation of $SL(2,\complex)$
modulo gauge transformations. If our previous arguments are correct, we
expect the dimension of the cohomology group $H^1(EndT)$ to be 12.
For the EH instanton this computation is not too difficult.
Recalling that the EH manifold is the smooth resolution of the algebraic
variety defined by the locus $W(x,y,z) = xy - z^2 = 0$ in $\complex^3$,
we see that the tangent bundle is an $SL(2,\complex)$ complex bundle whose
sections (vector fields), $\vec T=(T^x,T^y,T^z)$, satisfy the condition
$$
 yT^x+xT^y+2zT^z=0
\eqn\tangentbundle
$$
We now want to deform the tangent bundlle into a new complex bundle, defined by
$$
P_xT^x+P_yT^y+P_zT^z=0
\eqn\defortangent
$$
but in such a way so as to mantain the vanishing of the first Chern class.
The polynomial deformations of the tangent bundle $P_x,P_y,P_z$, which do not
modify the first Chern class, can be
identified as the polynomials in the variables $x,y,z$ of degree less or
equal to the degree of the components of the gradient to the locus.
In the case at hand we have $\vec \nabla W=(y,x,2z)$, so we need only
to consider polynomials of degree one
in $x,y,z$. Since each polynomial of degree one in three variables
is identified by four coefficients, we conclude that the cohomology
group $H^1(EndT)$ has dimension $3 \time 4 = 12$.
At the moment we are not able to relate the algebro-geometric basis to
the ``physical" basis, given by the gauge field zero--modes \azeromodes.
A proper understanding of this issue could help in determining the moduli
space of instanton connections on the EH manifold.

\chapter{Conclusions and Perpectives}
In this paper we have shown that an arbitrary conformally self--dual
gravitational background is solution of the classical equations of motion
of the effective supergravity theory arising from the heterotic string.
Specializing the gravitational background to the EH instanton,
we have computed all the relevant fermionic and bosonic zero--modes.

The interest of all these calculations, which we have hinted in the
Introduction, is in the fact that gravitational instantons may lead to SUSY
breaking via the formation of gravitino and gaugino condensates.

It is one of the miracles of supersymmetry that the computation of suitable
correlation functions, expanded around the saddle--point given by the
classical instantonic solution, leads to finite and space--time independent
results.
It turns out that the computation of the functional integral gets reduced
to the computation of a suitable combination of the zero--modes of the theory
integrated over the moduli space of the instanton, since the determinants
which are obtained from the functional integration over non--zero--modes
cancel by
supersymmetry. The first step in these calculations is thus the knowledge
of the zero--modes themselves, which is the subject we hope to have settled in
this work.
This computations were first performed
in [\rossi] for global SUSY theories and in [\konuno] for minimal $N=1$
supergravity. The present work should be the first of a series of papers
in which instanton calculus is hopefully extended to generic supergravity
theories, arising in the low energy limit of the heterotic string theory.

The deep reason why some correlation functions are constant
in certain theories, probably lies in their connection
with topological theories [\wittop]. From a phenomenological point of view,
theories with $N=1$ SUSY seem to be preferable with respect to those
with $N=2$\foot{A lot of progress has been recently made in the understanding
of global $N=2$ supersymmetric theories\REF\nequaldue{N.~Seiberg and E.~Witten,
hep-th/9407087, RU-94-52, IAS-94-43}[\nequaldue]}, but in the former case,
unfortunately, it has not been yet possible to relate the existence of
constant correlation functions to known
geometrical properties of the theory. Expecially in the case in which gravity
is present such an interpretation would be highly desirable, given also the
difficulty of defining in such theories, meaningful gauge invariant
quantities.
In a paper companion of this one we will discuss these matters together
with the form of the Ward identities appropriate for $N=1$ supergravity.

\ack
We would like to acknowledge fruitful discussions with L. Alvarez-Gaum\'e,
D. Anselmi, R. D'Auria, S. Ferrara, G. Gibbons, L. Girardello, C. Hull,
K. Konishi, A. Sagnotti, M. Testa, and G. Veneziano. Two of us
(M.B. and G.C.R.) would like to thank the organizers of the 1992 Triangular
Meeting at Crete, where their collaboration started.
\endpage

\Appendix{A}
In this Appendix we establish the conventions and notations used throughout
the paper.
Indices from the beginning of the latin alphabet $(a,b,c,\cdots)$ are frame
indices which range from $0$ to $3$ and are raised and lowered by the flat
Euclidean metric. Indices from the middle of the latin alphabet
$(i,j,k,\cdots)$ range from $1$ to $3$.
Spinor indices are usually denoted by letters
from the beginning of the greek alphabet ($\alpha,\beta,\gamma,\cdots$).
Indices from the middle of the greek alphabet
($\lambda,\mu,\nu,\cdots$) denote coordinate indices
which are raised and lowered by the metric tensor.

We fix the orientation of the volume measure by taking the permutation $1230$
as the fundamental one and defining $\varepsilon^{1230}=+1$.

For the Euclidean Dirac matrices we use a Weyl basis
$$
\gamma^a = \pmatrix{ 0 & \sigaaadot\cr
 \sigabdotb & 0 \cr}
\eqn\diracgamma
$$
where:
$$
\eqalign{
\sigma^0{}_{\aadot} &= -i \delta_{\aadot}\quad
\sigma^k{}_{\aadot} = \t^k{}_{\aadot} \cr
\bar\sigma_0{}^{\adota} &= +i \delta^{\adota} \quad
\bar\sigma_k{}^{\adota} = \t_k{}^{\adota}\cr}
\eqn\diracsigma
$$
with $\t^k\equiv \t_k$ the standard Pauli matrices.

Since the $\sigma^a$'s form a complete set, one has
$$
\sigaaadot \sigabdotb = 2 \delta_\alpha{}^\beta \delta_{\adot}{}^{\bdot}
\eqn\sigmacompletezza
$$

As usual we introduce the matrices
$$
\eqalign{
\sigma_{ab} & =\mezzo(\sigma_a \bar\sigma_b-\sigma_b \bar\sigma_a)\cr
\bar\sigma_{ab} & =\mezzo(\bar\sigma_a \sigma_b-\bar\sigma_b \sigma_a)\cr}
\eqn\defsigmaab
$$
which satisfy the relations
$$
\eqalign{
\sigaaadot \sigbadotb &= \delta^a{}_b \delta_\alpha{}^\beta
 +(\sigma^a{}_b)_\alpha{}^\beta\cr
\sigaadota \sigbabdot &=  \delta_a{}^b \delta^{\adot}{}_{\bdot} +
(\bar\sigma^b{}_a)^{\adot}{}_{\bdot} \cr}
\eqn\sigmadue
$$
The matrices \defsigmaab~are the generators of the Lorentz group
$SU(2)_L\otimes SU(2)_R$ and can be also rewritten as
$$
\eqalign{
(\sigma^a{}_b)_\alpha{}^\beta &= +i (\eta_k){}^a{}_b
(\t^k){}_\alpha{}^\beta\cr
({\bar\sigma}{}^a{}_b)^{\adot}{}_{\bdot} &= +i ({\bar\eta}_k)^a{}_b
(\t^k)^{\adot}{}_{\bdot}\cr}
\eqn\Lorentzsigma
$$
where (${\bar\eta}_k{}^{ab}$) $\eta_k{}^{ab}$ are the (anti)self--dual 't Hooft
symbols
$$
\eqalign{
\eta_k{}^{ab} &= \delta^a{}_k\delta^b{}_0 - \delta^a{}_0\delta^b{}_k
+ \varepsilon_k{}^{ij}
\delta^a{}_i\delta^b{}_j\cr
{\bar\eta}_k{}^{ab} &= - \delta^a{}_k\delta^b{}_0 + \delta^a{}_0
\delta^b{}_k +
\varepsilon_k{}^{ij} \delta^a{}_i\delta^b{}_j \cr}
\eqn\etathooft
$$
They satisfy the relations
$$
\eqalign{
\eta_i{}^{ab} \eta_{jbc} &= - \delta_{ij} \delta^a{}_c - \varepsilon_{ij}{}^k
(\eta_k)^a{}_c \cr
{\bar\eta}_i{}^{ab} {\bar\eta}_{jbc} &= - \delta_{ij} \delta^a{}_c -
\varepsilon_{ij}{}^k ({\bar\eta}_k)^a{}_c\cr
\eta_i{}^{ab} \eta^{icd} &= \delta^{ac} \delta^{bd} - \delta^{ad} \delta^{bc}
 + \varepsilon^{abcd} \cr
{\bar\eta}_i{}^{ab} {\bar\eta}^{icd} &= \delta^{ac} \delta^{bd} -
\delta^{ad} \delta^{bc} - \varepsilon^{abcd} \cr}
\eqn\etacontrac
$$
The duality properties of the 't Hooft symbols induce the following duality
properties of the Lorentz generators
$$
\eqalign{
\sigma^{ab} &= {1 \over 2} \varepsilon^{ab}{}_{cd} \sigma^{cd}, \cr
\bar\sigma^{ab} &= - {1 \over 2} \varepsilon^{ab}{}_{cd}{\bar\sigma}^{cd} \cr}
\eqn\sigmadual
$$
Spinor indices are raised and lowered by the symplectic metric \nextline
$$
\eqalign{
\eps &= i (\t_2)^{\alpha\beta} \quad \epsinv =-i (\t_2)_{\alpha\beta} \cr
\epsdot &= i (\t_2)^{\adot\bdot} \quad \epsdotinv = -i (\t_2)_{\adot\bdot} \cr}
\eqn\spinorepsilon
$$
which satisfy
$$
\eqalign{
\eps \varepsilon_{\beta\gamma} &= \delta^\alpha{}_\gamma\cr
\epsdot \varepsilon_{\bdot\dot\gamma} &= \delta^{\adot}{}_{\dot\gamma}\cr}
\eqn\epsilondue
$$
As a consequence, the matrices $\bar\sigma^a$ and $\sigma^a$ are related by
$$
\sigaadotb = - \eps \sigabbdot \epsdot
$$
and the Lorentz generators with two upper indices are symmetric
under the interchange of the spinor indices, \ie
$$
\eqalign{
\eps (\sigma_{ab})_\beta{}^\gamma &= (\sigma_{ab})^{\alpha\gamma} =
(\sigma_{ab})^{\gamma\alpha}\cr
\epsdotinv ({\bar\sigma}_{ab})^{\bdot}{}_{\dot\gamma} &=
({\bar\sigma}_{ab})_{\adot} {}^{\dot\gamma} =
(\bar\sigma_{ab})_{\dot\gamma}{}^{\adot}\cr}
\eqn\spinsymm
$$
In terms of the $\gamma$ matrices the duality properties of the Lorentz
generators translate into the equations
$$
\eqalign{
\gamma^{abcd} &= \varepsilon^{abcd} \gamma_5\cr
\gamma^{abc} &= \varepsilon^{abcd} \gamma_5 \gamma_d \cr
\gamma^{ab} &= -{1\over 2} \varepsilon^{abcd} \gamma_5 \gamma_{cd}\cr
\gamma^{a} &= -{1 \over 6} \varepsilon^{abcd} \gamma_5 \gamma_{bcd}\cr
\unita &= {1\over 24} \varepsilon^{abcd} \gamma_5 \gamma_{abcd} \cr}
\eqn\gammadual
$$

Right--handed (dotted) spinors, $\epsilon$, satisfy the equation
$\gamma_5 \epsilon = - \epsilon$. Left--handed (undotted) spinors, $\eta$,
satisfy the equation $\gamma_5 \eta =  \eta$.

Using the 't Hooft symbols, one can define the left-invariant one--forms
$$
\s_i = {1\over {r^2}} \eta_{i\mn} x^\mu dx^\nu
\eqn\cartesianLforms
$$
and the two--forms
$$
d\s_i = {1\over {r^2}} \eta_{i\mn} dx^\mu dx^\nu -
{2\over r^4} \eta_{i\mn} x_\r x^\m dx^\r dx^\n
\eqn\cartesiantwoforms
$$
where $r^2 = x^2 + y^2 + z^2 + t^2$. From \cartesianLforms~and
\cartesiantwoforms~one has $d\sigma_i =
\varepsilon_i{}^{jk}\sigma_j\wedge\sigma_k$.

In terms of the spherical coordinates
$$
\eqalign{
x + i y &= r \cos{\theta\over2} \exp{{i\over 2} (\psi + \varphi)}\cr
z + i t &= r \sin{\theta\over2} \exp {{i\over 2} (\psi - \varphi)}\cr}
\eqn\sphercoord
$$
the left-invariant one--forms, $\s_i$ have the explicit expression
$$
\eqalign{
\sigma_x &= {1\over 2} (\sin{\psi} d\theta -
 \cos{\psi} \sin{\theta} d\varphi)\cr
\sigma_y &= -{1\over 2} ( \cos{\psi} d\theta +
 \sin{\psi} \sin{\theta} d\varphi)\cr
\sigma_z &= {1\over 2} ( d\psi + \cos{\theta} d\varphi)\cr}
\eqn\leftforms
$$
Their dual vector fields are (see \killinguno, \killingdue)
$$
L_i = -{i\over 2} \eta_{i\mu}{}^\nu x^\mu \de_\nu
\equiv - i \xi _{(i)}^{\m} \de_{\m}
\eqn\cartesianLmomenta
$$
They satisfy the $SU(2)$ algebra: $[L_i,L_j] = i \varepsilon_{ij}{}^k L_k$.
In terms of spherical coordinates they read
$$
\eqalign{
L_1 &= -i (\sin{\psi} {\ddtheta} - {\cos{\psi}\over \sin{\theta}} {\ddphi}
+ {\cos{\theta} \cos{\psi} \over \sin{\theta}} {\ddpsi})\cr
L_2 &=  i ( \cos{\psi} {\ddtheta} + {\sin{\psi}\over \sin{\theta}} {\ddphi}
- {\cos{\theta} \sin{\psi} \over \sin{\theta}} {\ddpsi})\cr
L_3 &=  -i {\ddpsi} \cr}
\eqn\angularmomenta
$$
Frequently used combinations of the first two vectors are
$$
L_\pm = L_1 \pm i L_2 = e^{\pm i \psi}
   (\mp {\ddtheta} + {i \over \sin{\theta}} {\ddphi}
- i { \cos{\theta} \over \sin{\theta}} {\ddpsi})
\eqn\angularpm
$$
In an analogous way one can define the right-invariant one--forms:
$$
\bar\sigma_i = {1\over {r^2}} \bar\eta_{i\mn} x^\mu dx^\nu
$$
or in terms of Euler angles:
$$
\eqalign{
\bar\sigma_x &= {1\over 2} (\sin{\varphi} d\theta -
 \cos{\varphi} \sin{\theta} d\psi)\cr
\bar\sigma_y &= -{1\over 2} ( \cos{\varphi} d\theta +
 \sin{\varphi} \sin{\theta} d\psi)\cr
\bar\sigma_z &= {1\over 2} ( d\varphi + \cos{\theta} d\psi)\cr}
\eqn\rightforms
$$
Their dual vector fields are (see \killinguno)
$$
\bar L_i = - {i\over 2} \bar\eta_i{}^{\mn} x_\mu \de_\nu
\equiv - i \bar \xi _{(i)}^{\m} \de_{\m}
\eqn\cartesianLbarmomenta
$$
They satisfy the $SU(2)$ algebra: $[{\bar L}_i,{\bar L}_j] =
-i \varepsilon_{ij}^k {\bar L}_k$.
In terms of spherical coordinates they read
$$
\eqalign{
{\bar L}_1 &= -i (\sin{\varphi} {\ddtheta} -
     {\cos{\varphi}\over \sin{\theta}} {\ddpsi}
+ {\cos{\theta} \cos{\varphi} \over \sin{\theta}} {\ddphi})\cr
{\bar L}_2 &=  i ( \cos{\varphi} {\ddtheta} +
     {\sin{\varphi}\over \sin{\theta}} {\ddpsi}
- {\cos{\theta} \sin{\varphi} \over \sin{\theta}} {\ddphi})\cr
{\bar L}_3 &=  -i {\ddphi} \cr}
\eqn\barangularmom
$$
The raising and lowering operators become
$$
{\bar L}_\pm = {\bar L}_1 \pm i {\bar L}_2 = e^{\pm i \varphi}
   (\mp {\ddtheta} + {i \over \sin{\theta}} {\ddpsi}
- i { \cos{\theta} \over \sin{\theta}} {\ddphi})
\eqn\barangularpm
$$
The elements of the group $SU(2)$ may be parametrize in terms of the rotations
matrices
$$
D_R (\theta,\varphi,\psi) =
e^{i \psi T_3 } e^{i \theta T_2} e^{i \varphi T_3}
\eqn\rotationmatrix
$$
where the $T$'s are the $SU(2)$ generators in the representation
${\underline R}$. The matrix elements of the rotation matrices
$$
D^l_{n,m}(\theta,\varphi,\psi) = e^{i n \psi } e^{i m \varphi}
d^l_{n,m}(\theta)
= \langle \theta,\phi,\psi \vert l,m,n \rangle \quad |m|\leq l, |n|\leq l
\eqn\rotationvector
$$
yield a spin $l$ representations of the two $SU(2)$ groups defined above
and one has
$$
\eqalign{
L^2 \vert l,m,n \rangle &= l(l+1) \vert l,m,n \rangle \cr
L_3 \vert l,m,n \rangle &= n \vert l,m,n \rangle \cr
{\bar L}^2 \vert l,m,n \rangle &= l(l+1) \vert l,m,n \rangle \cr
{\bar L}_3 \vert l,m,n \rangle &= m \vert l,m,n \rangle \cr }
\eqn\representation
$$
The explicit expression of the matrices $d^l_{n,m}(\theta)$ are not needed
in this work except for the case $l=1$, for which we have
$$
d^1_{n,m}(\theta) = \pmatrix{
{1+\cos\theta \over 2} & {\sin\theta \over \sqrt{2} } &
{1-\cos\theta \over 2}\cr
-{\sin\theta\over \sqrt{2} } & \cos\theta & {\sin\theta\over \sqrt{2}}\cr
{1-\cos\theta \over 2} & -{\sin\theta \over \sqrt{2} } &
{1+\cos\theta \over 2}\cr}
\eqn\rotmat
$$
The matrix elements of the raising and lowering operators can be derived from
the equations
$$
\eqalign{
L_{\pm} \vert l,m,n \rangle &=
      \sqrt{(l \mp n)(l \pm n+1)} \vert l,m,n\pm 1 \rangle \cr
{\bar L}_\pm \vert l,m,n \rangle &=
      \sqrt{(l \mp m)(l \pm m+1)} \vert l,m \pm 1,n \rangle \cr}
\eqn\matrixelements
$$
\REF\harmonics{J.N.~Goldberg, A.J.~MacFarlane, E.T.~Newman, F.~Rohrlich
and C.G.~Sudarshan, J. Math. Phys. {\bf 8} (1967) 2155; \nextline
E.T.~Newman and R.~Penrose, J. Math. Phys.{\bf 7} (1966) 863;\nextline
E.D.~Fackerell and R.G.~Crossman, J. Math. Phys.{\bf 18} (1977) 1849.}
More details on rotation matrices and spinor weighted spherical harmonics
can be found in [\harmonics] or in standard textbooks of Quantum Mechanics.
\endpage

\Appendix{B}
\REF\kron{N.J.~Hitchin, Math.Proc.Camb.Phil.Soc. {\bf 85} (1979) 465;
\nextline P.B.~Kronheimer, J. Diff. Geom. {\bf 29} (1989) 665;
ibid. {\bf 29} (1989) 685.}
In this Appendix we recall some known facts about the geometry of ALE
manifolds with a special emphasis on those aspects that are more relevant
for the problems we are dealing with in this paper.

The nice thing about
ALE spaces is that they admit a complete topological classification,
the so--called $A-D-E$ classification, in terms of the kleinian subgroups
of $SU(2)$ [\kron]. Indeed to each one of such discrete subgroups,
$\Gamma$, one can associate an ALE manifold, $M$, which is
the smooth resolution of a certain singular variety in $\complex^3$
with boundary $\de M=S^3/\Gamma$.
Referring to the existing literature on the subject and
denoting by $\chi$ the Euler characteristic and by $\tau$ the
Hirzebruch signature, let's recall the relevant facts about
the $A-D-E$--classification. The classification is made in terms of
two discrete series corresponding to the Dinkin diagrams, $A_k$ and $D_k$,
and of the three exceptional cases corresponding to the exceptional groups
$E_6$, $E_7$ and $E_8$. The $k^{th}$ representative, $M_{A_k}$,
of the $A$--series corresponds to the smooth resolution
of the singular algebraic variety defined by the equation
$xy=z^{k+1}$ in $\complex^3$. Its boundary is the lens space $L(k+1,1)$
which coincides with $S^3$ modded out by the cyclic group $Z_{k+1}$,
\ie~$\de M_{A_k}=S^3/Z_{k+1}$. The manifold $M_{A_k}$ has $\chi=k+1$,
$\tau=-k$. The $k^{th}$ representative, $M_{D_k}$, of the $D$--series
corresponds to the resolution of the singular algebraic
variety defined by the equation $x^2z+y^2=z^{k-1}$ in $\complex^3$.
Its boundary is $S^3/D^*_k$ with $D^*_k$ the double-covering of the
dihedral group of order $k$. The manifold $M_{D_k}$ has $\chi=k+1$, $\tau=-k$.
The three ALE spaces associated to the Dinkin diagrams
of the exceptional groups can be described as follows.
The $E_6$ representative, $M_{E_6}$, corresponds to the resolution of the
variety $z^4=x^2+y^3$, its boundary is $S^3/T^*$ with $T^*$ the binary
tetrahedral group. $M_{E_6}$ has $\chi=7$, $\tau=-6$.
The $E_7$ representative, $M_{E_7}$, corresponds to the resolution of
the variety $yz^3=x^2+y^3$, its boundary is $S^3/O^*$ with $O^*$
the binary octahedral group. $M_{E_7}$ has $\chi=8$, $\tau=-7$.
The $E_8$ representative, $M_{E_8}$, corresponds to the resolution
of the variety $z^5=x^2+y^3$, its boundary is $S^3/I^*$ with $I^*$
the binary icosahedral group. $M_{E_8}$ has $\chi=9$, $\tau=-8$.
Barring geometrical subtleties, once the singularity has been smoothed out,
one can introduce on the resulting ALE manifold a Riemannian structure,
\ie~a metric.

An explicit class of self--dual metrics, which corresponds to the ALE manifolds
of the $A$--series, is given by the Gibbons--Hawking multi--center ansatz
$$
ds^2 = V^{-1}({\vec x}) (d\t + {\vec\omega}\cdot d {\vec x})^2 +
V({\vec x})d{\vec x}\cdot d {\vec x}
\eqn\multicenter
$$
with $V$ and $\vec\omega$ related by
$$
{\vec \nabla}V = {\vec \nabla}\times{\vec\omega}
\eqn\rotor
$$
to ensure the self--duality of the curvature.
Since the rigt--hand--side of eq.\rotor~is formally divergenceless up to
point-like sources, the most general form of $V$ is
$$
V({\vec x}) = \epsilon + 2m \sum_{i=1}^{ k+1} {1\over\mid {\vec x}-
{\vec x}_i\mid}
\eqn\potential
$$
with $\epsilon, m$ arbitrary parameters. Following the standard notations
used in this context, in the above formulae we have set the Newton
constant, $G_N$, equal to 1.
The choice of a common value of $m$
for all the sources is dictated by the requirement that the apparent
Dirac--string singularities induced in $\omega$ be removable coordinate
singularities of the metric.
The choice $\epsilon=0, m=\mezzo$ corresponds to an admissible metric
on the $k^{th}$ representative of the $A$--series space, $M_{A_k}$.
In particular the choice $k=0$ corresponds to a reparametrization of flat
four--dimensional
Euclidean space, while $k=1$ corresponds to the Eguchi--Hanson
istanton which is discussed at length in this paper.
The choice $\epsilon=1$ corresponds to the multi-Taub-NUT spaces which are
neither complex nor ALE spaces.

{}From the observation that ALE metrics are conjectured to make the
Einstein-Hilbert action non-negative definite, we conclude that
self--dual ALE metrics are absolute
minima of the action, since in each set of topologically
equivalent metrics, characterized by a common behavior at the boundary,
self--dual metrics have zero action.

\REF\gsvy{B.R.~Greene, A.~Shapere, C.~Vafa and S.-T.~Yau, Nucl.Phys.
{\bf B337} (1990) 1.}
For the sake of making contact with the complex geometry of supersymmetric
backgrounds which may allow consistent propagation of the
string, it is important to observe that ALE manifolds may be thought
of as non compact Calabi-Yau manifolds [\gsvy, \torinesi].
Indeed, thanks to the self--duality of the curvature,
the holonomy group of the ALE manifold with boundary $S^3 / \Gamma$
is an $SU(2) / \Gamma$ subgroup of
the Euclidean Lorentz group, $SO(4)\sim SU(2)_L\otimes SU(2)_R$. As a
consequence in these backgrounds there always exist two covariantly constant
right--handed spinors, $\bar\epsilon^{(r)}_{\adot}, r=1,2$, out of which one
can construct three covariantly constant complex structures
$$
J^i{}_{ab} =
\bar \epsilon^{(r)} \bar \s_{ab} \bar \epsilon^{(s)} \s^i{}_{rs}
\eqn\complexstructure
$$
which together with the identity, $J^0{}_{ab}=\d_{ab}$, satisfy the
quaternionic algebra. Choosing one of the complex structures, say $J^3$, as the
reference one, it is possible to introduce complex coordinates on the manifold
and associate to it a closed two--form of type (1,1),
$K=J^3_{ab}e^a{}_\m e^b{}_\n dx^\m dx^\n=K_{I\bar J}dz^I\wedge dz^{\bar J}$,
which will play the role of K\"ahler form.
This means that there exists a complex
coordinate system in which the hermitean metric is k\"ahlerian.
Actually ALE manifolds are hyperk\"ahlerian, due to the presence
of the two other covariantly constant complex
structures, $J^1$ and $J^2$.
The latter generate (anti)holomorphic two--forms of
type (2,0) (or (0,2)) which may be interpreted as the nowhere vanishing
(anti)holomorphic forms which characterize a Calabi-Yau
manifold of complex dimension two.
Since the only compact manifold of this class is known to be K3, one is led
to identify ALE manifolds as ``non-compact Calabi-Yau twofolds''.

The local polynomial ring of the algebraic equation which characterizes an
ALE manifold is the ring of deformations of its complex structure [\kron].
The number of generators of the ring for the $A$--series turns out to be
$k$ and coincides with the number of complex parameters in the admissible
deformations of the complex structure.
These parameters could be reabsorbed through non--analytic changes of
coordinates, in much the same way as one can eliminate
the explicit dependence on the moduli from the metric of a Riemann
surface. Another set of $k$ (real) parameters corresponds to the admissible
deformations of the K\"ahler structure. In the case of the
multi--center ansatz \multicenter--\potential, once the origin of
the three--dimensional coordinate system has been chosen
to coincide with one of the centers in the potential $V$,
the metric will depend on $3k$ parameters which can be made to coincide
with the $3k$ coordinates of the remaining centers.
These $3k$ parameters are in one to one correspondence with the zero--modes
of the metric, \ie~with the zero--modes of the Liechnerowicz operator in
the background given by the metric \multicenter~with the choice
$\epsilon=0, m=\mezzo$ in \potential.
The relevant cohomology group is $H^{(1,1)}(M_{A_k},\de M_{A_k})$, \ie~the
group of closed two--forms of Hodge type (1,1) which vanish
sufficiently fast on the boundary so as to have a finite norm.
For ALE manifolds this cohomology group consists only of (harmonic) self--dual
two--forms and has dimension $h_{(1,1)}=k$.
Combining these $k$ self--dual two--forms with the three covariantly constant
(and as such not normalizable) antiself--dual two--forms
which are always present on hyperk\"ahler manifolds, one gets $3h_{(1,1)}=3k$
deformations of the metric which can be shown to be zero--modes of the
Liechnerowicz operator.
Notice that for complex two-dimensional manifolds the relative cohomology group
$H^{(1,1)}(M_{A_k},\de M_{A_k})$
is responsible both for the complex deformations of the complex structure and
for the real deformations of the K\"ahler class.
The $k$ deformations of the K\"ahler structure can be
``complexified'' by taking as imaginary parts the $k$ self--dual two--forms,
of type (1,1). These two--forms  can be interpreted as zero--mode
fluctuations of the antisymmetric tensor around the corresponding
ALE background.

The last topic we would like to briefly discuss in this Appendix
is the deformation of the (vacuum) gauge bundle [\wittdue, \wittendt].
Through the ``standard embedding'' the gauge bundle is identified with the
tangent bundle to the ALE manifold.
Since for a K\"ahler manifold the curvature two--forms have only non--vanishing
components of type (1,1), the transition functions may be chosen to be
holomorphic. The tangent bundle, $T_M$ or more simply $T$, turns out
to be a holomorphic vector bundle. Thanks to the vanishing of the Ricci scalar,
$T$ has
vanishing first Chern class. Writing the algebraic equation of the singular
variety in $\complex^3$ associated to the ALE manifold in the implicit
form, $W(x,y,z)=0$, the tangent vectors $\vec t$, satisfy
$$
t^x{\de W\over \de x} + t^y{\de W\over \de y} + t^z{\de W\over \de z}= 0
\eqn\tangent
$$
In much the same way as an algebraic singularity can be deformed to a
``nearby" (non--singular) variety through the elements of its polynomial ring,
the defining equation of the tangent bundle \tangent~can be modified to
a ``nearby" equation with polynomial coefficients defining a new vector bundle
$V$.
The new holomorphic vector bundle $V$ may be equivalent to $T$ in the
topological sense, but in general it will not be equivalent to $T$ in the
holomorphic sense.
In order to shift the initial solution of the heterotic string equations of
motion to a ``nearby" solution, admissible gauge bundle $V$ are naturally
related to
holomorphic deformations of $T$ which preserve the rank
and the vanishing of the first Chern class.
Somehow in analogy with the deformations of tangent bundles to
hypersurfaces embedded in complex projective spaces [\wittdue, \wittendt], the
relevant deformations of \tangent~should be
in one to one correspondence with
the polynomials in $x,y,z$ of degree less or equal to the smallest among the
degrees of ${\de W\over \de x}$,${\de W\over \de y}$ and ${\de W\over \de z}$.
In view of the identification between the gauge connection
and the connection on the tangent bundle, these deformations should
correspond to zero--mode fluctuations of the gauge field
around the considered background.
In two complex dimensions self--duality and holomorphicity are tightly bound.
In complex coordinates ${z^I, I=1,2}$ a hermitean self--dual connection, $A$,
has components:
$$
\eqalign{
A_I &= U^{-1} \de_I U \cr
A_{\bar I} &= (\de_{\bar I} {\bar U}) {\bar U}^{-1} \cr}
\eqn\holoconn
$$
with $U \neq {\bar U}$. The connection \holoconn~is not a pure gauge.
The components of Hodge type (2,0) and (0,2) of the
field--strength $F$ vanish, but the components of type (1,1) do not.
Deforming the tangent bundle to a new vector bundle with self--dual curvature
(field--strength) is tantamount to find a new gauge connection
such that the above properties of $F$ are mantained. This means that
the zero--modes of the gauge connection are in one to one correspondence
with holomorphic one--forms in the adjoint representation of the structure
group. The structure group is $SL(2,\complex)$,
which is the complexification of the $SU(2)$ group where the
gauge istanton lives. The relevant deformations are those belonging to the
cohomology group $H^1(End(T))$, with $End(T)$ the bundle of
endomorphisms of the tangent bundle, $T$ [\wittendt, \wittdue].

\endpage
\refout
\end